\RequirePackage{fix-cm}
\documentclass[conference]{IEEEtran}
\IEEEoverridecommandlockouts
\usepackage{amsmath,amssymb,amsfonts}
\usepackage{algorithmic}
\usepackage{graphicx}
\usepackage{textcomp}

\usepackage[utf8]{inputenc}

\usepackage{enumitem}
\setlistdepth{9}
\newlist{proofItemize}{itemize}{9}
\setlist[proofItemize]{leftmargin=2mm}

\usepackage{hyperref}
\usepackage[numbers]{natbib}
\usepackage[nolist,nohyperlinks]{acronym}

\usepackage{amsmath,amssymb}
\usepackage{mathtools}
\usepackage{multirow}

\usepackage{listings}

\usepackage{proof, xspace}
\usepackage[dvipsnames,table]{xcolor}

\usepackage{tikz}
\usetikzlibrary{fit, positioning, calc, arrows, arrows.meta,
  decorations.pathreplacing, decorations.markings, decorations.text,
  shapes, tikzmark, backgrounds, patterns}

\usepackage{fancyvrb}
\usepackage{unison}

\usepackage{xspace}
\usepackage[justification=centering]{subfig}

\usepackage{tcolorbox}
\tcbuselibrary{listings, skins}

\usepackage[textsize=footnotesize]{todonotes}

\usepackage{colortbl}
\newcolumntype{g}{>{\columncolor{Gray!40}}r}
\usepackage{hhline}

\usepackage{url}

\usepackage{amsthm}
\newtheorem{theorem}{Theorem}
\newtheorem{definition}{Definition}
\begin{acronym}
\acro{SecCG}{Secure by Construction Code Generation}
\acro{ASLR}{Address Space Layout Randomization}
\acro{DPA}{Differential Power Analysis}
\acro{IR}{Intermediate Representation}
\acro{MIR}{Machine Intermediate Representation}
\acro{ESV}{Extract Shared Variables}
\acro{DMG}{Datalog Model Generation}
\acro{LRA}{LLVM Register Allocation}
\acro{CP}{Constraint Programming}
\acro{COP}{Constraint Optimization Problem}
\acro{LNS}{Large Neighborhood Search}
\acro{HD}{Hamming Distance}
\acro{HW}{Hamming Weight}
\acro{MCR}{Multicompiler}
\acro{SMT}{Satisfiability Modulo Theory}
\acro{PSC}{Power Side Channel}
\acro{LNS}{Large Neighborhood Search}
\acro{TBL}{Transition-Based Leakage}
\acro{VBL}{Value-Based Leakage}
\acro{ROT}{Register-Overwrite Transition}
\acro{MOT}{Memory-Overwrite Transition}
\acro{MRE}{Memory-Remnant Effect}
\acro{RNL}{Register Neighbor Leakage}
\acro{IPI}{Instruction-Pair Interaction}
\acro{OT}{Other Transitions}
\acro{ME}{Middle-End}
\acro{FE}{Front-End}
\acro{BE}{Back-End}
\acro{TSC}{Timing Side Channel}
\acro{FI}{Fault Injection}
\acro{CT}{Constant Time}
\acro{MS}{Memory Safety}
\acro{IFL}{Information-Flow Leakage}
\acro{RS}{Residual Program State}
\acro{DSL}{Domain Specific Language}
\acro{ML}{Mitigation Level}
\acro{ISA}{Instruction Set Architecture}
\end{acronym}

\newcommand{\andrule}      {~~~}
\newcommand{\andcons}      {~\land~}

\newcommand{\infr} [3] []  {\infer[\textsc{#1}]{#3}{#2}}

\newcommand{\codebackcolor}{gray!10!white}
\newcommand{\colbackcolor}{gray!5!white}
\newcommand{\colframecolor}{white!60!black}
\newcommand{\colbackts}{gray!5!white}
\newcommand{\colframets}{white!60!black}

\definecolor{keywordCommentColor}{rgb}{0.090000, 0.55, 0.20}
\definecolor{stringColor}{rgb}{0.558215, 0.000000, 0.135316}
\definecolor{typeColor}{rgb}{0.6, 0.000000, 0.3}
\definecolor{localColor}{rgb}{0.6, 0.000000, 0.3}
\definecolor{ndkeywordColor}{rgb}{0.0, 0.558215, 0.558215}
\definecolor{commentsColor}{rgb}{0.0, 0.558215, 0.558215}
\definecolor{keywordColor}{rgb}{0.000000, 0.000000, 0.635294}
\definecolor{newgray}{rgb}{0.3, 0.3, 0.3}
\definecolor{agreen}{rgb}{0.0, 0.36, 0.15}
\definecolor{rgreen}{rgb}{0.13, 0.26, 0.12}

\newcommand{\basicCodeStyle}{\ttfamily\footnotesize\color{newgray}}

\lstdefinestyle{cstylepsc}{%
  language=C,
  frame=tb,
  numberblanklines=false,
  escapeinside=@@,
  basicstyle=\basicCodeStyle,
  framextopmargin=-10pt,
  morekeywords={u32},
  numbers=left,
  xleftmargin=15pt, %
  keywordstyle=\color{keywordColor}\bfseries,
  ndkeywordstyle=\color{ndkeywordColor}\bfseries,
  identifierstyle=\color{black}\ttfamily,
  commentstyle=\itshape\ttfamily\textcolor{commentsColor},
  stringstyle=\color{stringColor}\ttfamily,
  morekeywords = [2]{pub, Public},
  keywordstyle = [2]\color{green!40!black}\itshape,
  morekeywords = [3]{key, Secret},       
  keywordstyle = [3]\color{stringColor}\itshape,
  morekeywords = [4]{mask, Random},       
  keywordstyle = [4]\itshape,
}

\lstdefinestyle{modelingstyle}{%
        basicstyle = \ttfamily\small,
        mathescape = true,
        keywords={def, pred, exists, forall, ite, return, in, sum,
        sorted, then, else, if, constraint, max, branch},
        keywordstyle=\color{keywordColor}\bfseries,
}

\lstdefinestyle{typedunistyle}{%
        basicstyle = \ttfamily\small,
        mathescape = true,
        keywords={xor, copy, in, out},
        keywordstyle=\color{keywordColor}\bfseries,
        morekeywords = [2]{pub, t0, t3, Public},
        keywordstyle = [2]\color{green!40!black}\ttfamily\bfseries,
        morekeywords = [3]{key,  t2, t5, Secret},       
        keywordstyle = [3]\color{stringColor}\ttfamily\bfseries,
        morekeywords = [4]{mask,t1, t4, t6, t7, t8, t9, t10, Random},
        keywordstyle = [4]\color{brown!40!black}\ttfamily\bfseries,
}

\lstdefinestyle{unistyle}{%
        basicstyle = \ttfamily\small,
        mathescape = true,
        keywords={xor, copy, in, out},
        keywordstyle=\color{keywordColor}\bfseries,
        morekeywords = [2]{pub, Public},
        keywordstyle = [2]\color{green!60!blue!80!black}\ttfamily\bfseries, % \color{green!40!black}\ttfamily\bfseries,
        morekeywords = [3]{key, Secret},       
        keywordstyle = [3]\color{stringColor}\ttfamily\bfseries,
        morekeywords = [4]{mask, Random},
        keywordstyle = [4]\color{brown!40!black}\ttfamily\bfseries,
}

\lstdefinestyle{cstyle}{%
        language=C,
        basicstyle = \ttfamily\small,
        mathescape = true,
        morekeywords = {u32},
        keywordstyle=\color{keywordColor}\bfseries,
        numbers=left,
        xleftmargin=15pt, % 
        identifierstyle=\color{black}\ttfamily,
        commentstyle=\itshape\ttfamily\textcolor{commentsColor},
        morekeywords = [2]{pub,b},
        keywordstyle = [2]\color{green!60!blue!80!black}\ttfamily\bfseries, % \color{green!40!black}\ttfamily\bfseries,
        morekeywords = [3]{key,a},       
        keywordstyle = [3]\color{stringColor}\ttfamily\bfseries,
        morekeywords = [4]{mask,a0,b0,r01},
        keywordstyle = [4]\color{brown!40!black}\ttfamily\bfseries,
}

\lstdefinestyle{mipsstyle}{%
        comment = [l]{\#},
        escapeinside={!!},
        frame=none,
        keywordstyle=\color{keywordColor}\bfseries\ttfamily\em,
        identifierstyle=\color{black}\ttfamily,
        commentstyle=\itshape\ttfamily\textcolor{commentsColor},
        stringstyle=\color{Mahogany}\ttfamily,
        basicstyle=\ttfamily\small,%% \basicCodeStyle,
        keywords={, xor, xori, andi, b, lbu, lui, addu, lw, sw, add, addi,
        addiu, jr, jal, nop, move, slti, beqz, blez, mul, },
        morekeywords = [2]{\$zero,\$sp,\$ra,\$a0,\$a1,\$a2,\$a3,\$v0,\$v1,
        \$t0,\$t1,\$t2,\$t3,\$t4,\$t5,\$t6,\$t7,\$t8,
        \$t9,\$s1,\$s2,\$s3,\$s4,\$s5, \$fp,\$gp},
        keywordstyle = [2]\color{keywordCommentColor}\bfseries, 
        numbers=left,
        xleftmargin=.08\textwidth,
}

\lstdefinestyle{armstyle}{%
        comment = [l]{\@},
        escapeinside={!!},
        frame=none,
        keywordstyle=\color{keywordColor}\bfseries\ttfamily\em,
        identifierstyle=\color{black}\ttfamily,
        commentstyle=\itshape\ttfamily\textcolor{commentsColor},
        stringstyle=\color{Mahogany}\ttfamily,
        basicstyle=\ttfamily\small,%% \basicCodeStyle,
        keywords={, eor, eors, bx, str, ldr, },
        morekeywords = [2]{lr,r0,r1,r2,r3,r12,sp},
        keywordstyle = [2]\color{keywordCommentColor}\bfseries, 
        numbers=left,
        xleftmargin=.08\textwidth,
}

\DeclareFontFamily{U}{cmtex}{}
\DeclareFontShape{U}{cmtex}{m}{n}{
 <-> cmtex10
}{}
\newcommand{\cmtex}[1]{{%
  \ttfamily\makebox[0.5em]{%
    \usefont{U}{cmtex}{m}{n}\symbol{#1}%
  }%
}}
\newcommand{\ttoplus}{\cmtex{'015}}

\newcommand{\largestoh}{7\%\xspace}
\newcommand{\lowestsu}{75\%\xspace}
\newcommand{\largestsu}{8\xspace}

\newcommand{\lowestsuthumb}{2.2\xspace}
\newcommand{\largestsuthumb}{5.8\xspace}
\newcommand{\lowestsuthumbp}{P9\xspace}
\newcommand{\largestsuthumbp}{P1\xspace}

\newcommand{\lowestsumips}{75\%\xspace}
\newcommand{\largestsumips}{8.25\xspace}

\newcommand{\soverhead}{Oh\xspace}
\newcommand{\sspeedup}{Su\xspace}
\newcommand{\sslowdown}{Sd\xspace}

\newcommand{\mipsmaxcoh}{200K\xspace}
\newcommand{\thumbmaxcoh}{600K\xspace}

\newcommand{\mipsmaxcohunison}{57.3\xspace}
\newcommand{\thumbmaxcohunison}{2.5\xspace}

\newcommand{\mips}{Mips32\xspace}
\newcommand{\thumb}{ARM Thumb\xspace}

\newcommand{\toolname}{SecCG}
\newcommand{\toolnameshort}{SCG}

\def\halfcheckmark{\tikz\draw[scale=0.4,fill=red, draw=red](0,.35) -- (.25,0) -- (1,.7) -- (.25,.15) -- cycle (0.75,0.2) -- (0.77,0.2)  -- (0.6,0.7) -- cycle;}
\def\fullcheckmark{\tikz\draw[scale=0.4,fill=green!60!black, draw=green!60!black](0,.35) -- (.25,0) -- (1,.7) -- (.25,.15) -- cycle ;}
\def\cross{\tikz\draw[scale=0.4,fill=red, draw=red](.25,0) -- (1,.7) -- (.25,.15) -- cycle (0.75,0.1) -- (0.77,0.1)  -- (0.4,0.7) -- cycle;}

\usepackage{fontawesome}

\newcommand{\faast}{{\tiny\faAsterisk}\xspace}

\usepackage{comment}

\begin{document}
\title{Securing Optimized Code Against Power Side Channels}
%% \title{Conference Paper Title*\\
%% {\footnotesize \textsuperscript{*}Note: Sub-titles are not captured in Xplore and
%% should not be used}
%% \thanks{Identify applicable funding agency here. If none, delete this.}
%% }

\makeatletter
\newcommand{\linebreakand}{%
  \end{@IEEEauthorhalign}
  \hfill\mbox{}\par
  \mbox{}\hfill\begin{@IEEEauthorhalign}
}
\makeatother

\author{\IEEEauthorblockN{Rodothea Myrsini Tsoupidi}
\IEEEauthorblockA{\textit{Royal Institute of Technology KTH}\\
Stockholm, Sweden \\
tsoupidi@kth.se}
\and
\IEEEauthorblockN{Roberto Casta\~neda Lozano}
\IEEEauthorblockA{\textit{Independent Researcher}\\
Stockholm, Sweden \\
rcas@acm.org}
\and
\IEEEauthorblockN{Elena Troubitsyna}
\IEEEauthorblockA{\textit{Royal Institute of Technology KTH}\\
Stockholm, Sweden \\
elenatro@kth.se}
\linebreakand
\IEEEauthorblockN{Panagiotis Papadimitratos}
\IEEEauthorblockA{\textit{Royal Institute of Technology KTH}\\
Stockholm, Sweden \\
papadim@kth.se}
}

\maketitle

\begin{abstract}
  Side-channel attacks impose a serious threat to cryptographic
  algorithms, including widely employed ones, such as AES and RSA.
  These attacks take advantage of the algorithm implementation in
  hardware or software to extract secret information via side
  channels.
  Software masking is a mitigation approach against power
  side-channel attacks aiming at hiding the secret-revealing
  dependencies from the power footprint of a vulnerable
  implementation.
  However, this type of software mitigation often depends on
  general-purpose compilers, which do not preserve non-functional
  properties.
  Moreover, microarchitectural features, such as the memory bus and
  register reuse, may also leak secret information.
  These abstractions are not visible at the high-level
  implementation of the program.
  Instead, they are decided at compile time.
  To remedy these problems, security engineers often sacrifice code
  efficiency by turning off compiler optimization and/or performing
  local, post-compilation transformations.
  This paper proposes \ac{\toolname}, a constraint-based compiler
  approach that generates optimized yet secure against power
  side channels code.
  \ac{\toolname} controls the quality of the mitigated program by
  efficiently searching the best possible low-level implementation
  according to a processor cost model.
  In our experiments with twelve masked cryptographic functions up
  to 100 lines of code on \mips and \thumb, \ac{\toolname} speeds up
  the generated code from \lowestsu to \largestsu times compared to
  non-optimized secure code with an overhead of up to \largestoh
  compared to non-secure optimized code at the expense of a high
  compilation cost.
  For security and compiler researchers, this paper proposes a
  formal model to generate power side channel free low-level code.
  For software engineers, \ac{\toolname} provides a practical
  approach to optimize performance critical and vulnerable cryptographic
  implementations that preserves security properties against power
  side channels.
\end{abstract}


\section{Introduction}
\label{sec:intro}
% TODO: cite dsilva_correctness-security_2015-1 maybe

Cryptographic algorithms, symmetric/shared key or asymmetric/private
key ones, rely on safeguarding the shared secret key or the private
key, respectively.
%% Cryptographic algorithms such as AES and RSA, use secret keys to
%% encrypt or decrypt sensitive data.
%
The exposure of these keys to unintended users compromises the
security of these algorithms.
Unfortunately, the software implementation of cryptographic algorithms
may reveal information about their secret/private
keys~\cite{kocher_timing_1996-2}.
In particular, the attacker may observe what is termed
\emph{side-channel information}, notably observing the execution
time~\cite{kocher_timing_1996-2} or the power
consumption~\cite{kocher_differential_1999,joye_second-order_2005},
during the execution of the algorithm to extract information about the
secret keys.
These attacks are attractive especially because usually they do not
require expensive equipment.
This paper focuses on \ac{PSC} attacks.

Software masking is a widely-used approach to mitigate \ac{PSC}
attacks~\cite{rivain_provably_2010,coron_higher-order_2014}, hiding
secret information by splitting a secret into $n$ randomized shares.
The attacker has to retrieve all shares in order to acquire the secret
value.
While software masking can be an effective mitigation, compiler code
generation may optimize it away.
Moreover, \ac{TBL} sources, such as register reuse or memory-access order,
are decided at compile time by low-level compiler
transformations~\cite{bayrak_sleuth_2013,wang_mitigating_2019-1,eldib_formal_2014}.

To mitigate these compiler-induced power side-channel leaks at the
binary level there are techniques based on
compilation~\cite{wang_mitigating_2019-1,barthe_secure_2021,borrello_constantine_2021}
and binary rewriting with hardware
emulation~\cite{shelton_rosita_2021-1,veshchikov_use_2017-1,papagiannopoulos_mind_2017}.
All these approaches mitigate compiler-generated leakages using local
transformations~\cite{papagiannopoulos_mind_2017,wang_mitigating_2019-1,shelton_rosita_2021-1}.
The methods that depend on hardware emulation are typically accurate
but may introduce significant overhead~\cite{shelton_rosita_2021-1}
and are hardware specific.
For example, Rosita~\cite{shelton_rosita_2021-1}, an emulation-based
approach, propose a mitigation that introduces an overhead ranging
from 21\% to 64\% for ARM Cotrex M0.
\citeauthor{wang_mitigating_2019-1}~\cite{wang_mitigating_2019-1}
perform their mitigation using a standard compiler with no high-level
optimizations (-O0).
This is a common practice for security research to ensure the absence
of compiler-induced mitigation
invalidation~\cite{bayrak_sleuth_2013,vu_secure_2020}.
However, unoptimized code is highly inefficient, and
may even introduce additional leaks due to the
heavy use of the program stack, as discussed in
Section~\ref{sec:motivation}.

\citeauthor{vu_secure_2020} present an approach that enables
secure optimization of masked code at a higher
level~\cite{vu_secure_2020,vu_reconciling_2021-1}.
This approach applies high-level compiler optimizations by disallowing
secure-code removal and operand reordering (due to associativity of
some operations) and are able to generate correctly masked code.
However, they do not deal with \acp{TBL}.

Currently, the state-of-the-art approaches are unable to generate code
that is both efficient and secure in the face of \acp{TBL} that enable
\ac{PSC} attacks.
%% To summarize, current approaches to secure compilation for side-channel attacks
%% generate code that is either secure (does not leak secrets due to transitional
%% effects) or efficient, but not both. % \roberto{is this fair to say, or too bold?}
%
To address this challenge, this paper proposes \acf{\toolname}, an
optimizing compiler approach that provably preserves security
properties against \ac{PSC}.
At the middle-end, \ac{\toolname} handles code generated using
\textit{register promotion} (promoting program variables from memory to
registers) as a high-level optimization.
Then, \ac{\toolname} uses a constraint-based method to generate code
that is secure against \ac{PSC} attacks.
\ac{\toolname} controls the quality of the mitigated program by
efficiently searching the best possible low-level implementation
according to a processor cost model~\cite{lozano_combinatorial_2019}.
The security model of \ac{\toolname} is hardware agnostic and can be
extended with additional architectural constraints.
\ac{\toolname} is suitable for predictable architectures with no
advanced microarchitectural features, such as caches or speculative
execution.
In our experiments with twelve masked implementations on \mips and
\thumb, \ac{\toolname} improves the execution time of the generated
code from \lowestsu to a speedup of \largestsu compared to
non-optimized code at a overhead of up to \largestoh compared to
non-secure optimized code.
This comes at a cost on compilation time and reduced scalability.
In summary, this paper makes the following contributions:
\begin{itemize}
\item a compiler approach to generate leak-free, low-overhead assembly
  code for high-level software-masked programs;
\item a constraint model for optimized and \ac{PSC}-secure code
  generation;
\item a proof that the constraint model guarantees the generation of
  secure code for a non-trivial leakage model; and
\item experimental results on two architectures showing that the
  performance overhead of our mitigation is low and its efficiency
  benefits are significant, compared to current approaches.
\end{itemize}

\section{Motivating Example}
\label{sec:motivation}

To motivate our approach, let us consider an example of a first-order
masked implementation.
First-order masking splits a secret value \texttt{k} into two shares,
(\texttt{m}, \texttt{mk}), where \texttt{m} is a uniformly distributed
random variable sampled at every execution of the algorithm;
\texttt{mk} = $\texttt{m}\oplus\texttt{k}$ is also uniformly
distributed ($\oplus$ denotes the exclusive OR operation).
Figure~\ref{lst:masked} shows a first-order masked C implementation of
exclusive OR, where \texttt{key} is a secret value (red),
\texttt{mask} is a uniformly random variable (brown), and \texttt{pub}
is a non-secret value (green).
At line 2, the algorithm creates the second share, \texttt{mk}, and at
line 3, it performs the exclusive OR operation with the
secret-independent value, \texttt{pub}.
At a high-level, the code of Figure~\ref{lst:masked} is secure against
power side channels but a binary implementation generated by a
standard, security-unaware compiler may leak information about
\texttt{key}.
%
%% For example, the random share \texttt{mk} may reuse the register of the
%% \texttt{mask} variable, which results in a value transition of the reused
%% register from \texttt{mask} to \texttt{mk}.
%% %
%% This transition may be observable in the power trace of a device and
%% leak information about the secret \texttt{key}.
%% %
%% Moreover, memory operations that use the same bus may also
%% reveal secret information~\cite{shelton_rosita_2021-1}.
For example, hardware-register reuse and memory-bus access order may
reveal secret
information~\cite{wang_mitigating_2019-1,shelton_rosita_2021-1,
  bayrak_sleuth_2013,eldib_formal_2014}.
These \acp{TBL} are a result of transitional effects, i.e., the power
effect of bits switching between one and zero and vice versa.

Figure~\ref{fig:xor-llvm} shows the \thumb assembly code
generated by the standard compiler LLVM~\cite{lattner_llvm_2004} for
the C code in Figure~\ref{lst:masked}.
The first three \texttt{str} instructions store the function arguments
that reside in registers \texttt{r0-r2} to the stack (lines 3-5).
Line 6 loads (\texttt{ldr}) the value of \texttt{rand} from the stack into
register \texttt{r1}.
Line 7 performs the first exclusive OR (line 2 in
Figure~\ref{lst:masked}) between registers \texttt{r1} and \texttt{r2}
(\texttt{key}) and stores the result in register \texttt{r1}.
Here, there is a transition for register \texttt{r1} from value
\texttt{mask} to \texttt{mk}, which leaks the secret \texttt{key}
(marked code at line~7).
Line 8 stores the content of \texttt{r1} to the stack and the value of
the memory bus that contains the \texttt{mask} at line 6 transitions
to \texttt{mk}.
This leads to another leak due to the transitional effect in the
memory bus (marked code at lines~6 and 8).
The rest of the code performs the second exclusive OR (line
10) and stores the final result on the stack (line 11).

\begin{figure*}
  \centering
  %% Figure C code
\begin{minipage}{0.47\textwidth}
  \begin{tcolorbox}[colback=\codebackcolor, colframe=\codebackcolor,
      top=-5pt, bottom = -5pt,
      right = 0pt, left = 0pt]
\begin{lstlisting}[style=cstyle]
u32 Xor(u32 pub, u32 mask, u32 key) {
  u32 mk = mask ^ key;
  u32 t = pub ^ mk;
  return t;
}
\end{lstlisting}
%% \begin{lstlisting}[style = cstyle]
%% u32 Xor(u32 pub, u32 mask, u32 key) {
%%   u32 mk = mask ^ key;
%%   u32 t = pub ^ mk;
%%   return t;
%% }
%% \end{lstlisting}
  \end{tcolorbox}
  \caption{\label{lst:masked} Masked exclusive OR implementation in C}
  \end{minipage}
\hfill
%% Figure opt
\begin{minipage}{0.47\textwidth}
  \begin{tcolorbox}[colback=\codebackcolor, colframe=\codebackcolor,
      top=-5pt, bottom = -25pt,
      right = 0pt, left = 5pt]
  \subfloat[Insecure (LLVM)\label{fig:xor-mem2reg-llvm}]{
    \begin{minipage}[b]{0.45\textwidth}
\begin{lstlisting}[style=armstyle,firstline=28,lastline=31]
@ r0: pub, r1: mask, r2: key
  !\tikzmark{opteors1}!eors	r1, r2!\tikzmark{opteors2}!
  eors	r0, r1
  ...
\end{lstlisting}
      \begin{tikzpicture}[remember picture,overlay]
\draw[fill=orange, rounded corners, opacity=0.3]
  ([shift={(-3pt,1.5ex)}]pic cs:opteors1) 
  rectangle 
  ([shift={(9pt,-0.65ex)}]pic cs:opteors2);
\end{tikzpicture}
      \vspace{-5pt}
    \end{minipage}
  }
  \subfloat[Secure (\toolname)\label{fig:xor-mem2reg-sec}]{
    \begin{minipage}[b]{0.45\textwidth}
\begin{lstlisting}[style=armstyle,firstnumber=2,firstline=29,lastline=31]
  !\tikzmark{optseors1}!eors	r2, r1!\tikzmark{optseors2}!
  eors	r0, r2
  ...
\end{lstlisting}
      \begin{tikzpicture}[remember picture,overlay]
\draw[fill=orange, rounded corners, opacity=0.3]
  ([shift={(-3pt,1.5ex)}]pic cs:optseors1) 
  rectangle 
  ([shift={(9pt,-0.65ex)}]pic cs:optseors2);
\end{tikzpicture}
      \vspace{-5pt}
    \end{minipage}
  }
  \end{tcolorbox}
  \vspace{15pt}
  \caption{\label{fig:xor-llvm-memreg} Compilation of function
    \texttt{Xor} applying register promotion}
\end{minipage}
%% Figure non-opt
\subfloat[Insecure (LLVM)\label{fig:xor-llvm}]{
    \begin{minipage}{0.5\textwidth}
  \begin{tcolorbox}[colback=\codebackcolor, colframe=\codebackcolor,
      top=-5pt, bottom = -5pt,
      right = 0pt, left = 0pt]
\begin{lstlisting}[style=armstyle,firstline=29,lastline=40]
@ r0: pub, r1: mask, r2: key
...
str	r0, [sp, #16] @ mem: pub
str	r1, [sp, #12] @ mem: rand
str	r2, [sp, #8]  @ mem: key
!\tikzmark{ldr1}!ldr	r1, [sp, #12]!\tikzmark{ldr2}! @ mem: rand
!\tikzmark{eors1}!eors	r1, r2!\tikzmark{eors2}!        @ proc: rand <- rand ^ key 
!\tikzmark{str1}!str	r1, [sp, #4]!\tikzmark{str2}!  @ mem: rand ^ key
ldr	r0, [sp, #16] @ mem: pub
eors	r0, r1        @ proc: pub <- pub ^ rand ^ key
str	r0, [sp]      @ mem: pub ^ rand ^ key
...
\end{lstlisting}%
    \begin{tikzpicture}[remember picture,overlay]
\draw[fill=orange, rounded corners, opacity=0.3]
  ([shift={(-3pt,1.5ex)}]pic cs:eors1) 
  rectangle 
  ([shift={(9pt,-0.65ex)}]pic cs:eors2);

\draw[fill=yellow, rounded corners, opacity=0.3]
  ([shift={(-3pt,1.5ex)}]pic cs:ldr1) 
  rectangle 
  ([shift={(6pt,-0.60ex)}]pic cs:ldr2);

\draw[fill=yellow, rounded corners, opacity=0.3]
  ([shift={(-3pt,1.5ex)}]pic cs:str1) 
  rectangle 
  ([shift={(9pt,-0.65ex)}]pic cs:str2);
\end{tikzpicture}
    \vspace{-8pt}
  \end{tcolorbox}
  \end{minipage}
  }
  \subfloat[Secure (\toolname with no register promotion)\label{fig:xor-sec}]{
    \begin{minipage}{0.5\textwidth}
\begin{tcolorbox}[colback=\codebackcolor, colframe=\codebackcolor,
      top=-5pt, bottom = -5pt,
      right = 0pt, left = 0pt]
\begin{lstlisting}[style=armstyle,firstline=29,lastline=40]
@ r0: pub, r1: mask, r2: key
...
str	r1, [sp, #12] @ mem: rand
ldr	r1, [sp, #12] @ mem: rand
str	r0, [sp, #16] @ mem: pub
!\tikzmark{sstr1}!str	r2, [sp, #8]!\tikzmark{sstr2}!  @ mem: key
!\tikzmark{seors1}!eors	r2, r1!\tikzmark{seors2}!        @ proc: key <- sec ^ rand
!\tikzmark{sstr3}!str	r2, [sp, #4]!\tikzmark{sstr4}!  @ mem: key ^ rand
ldr	r0, [sp, #16] @ mem: pub
eors	r0, r2        @ proc: pub <- pub ^ key ^ rand
str	r0, [sp]      @ mem: pub ^ rand ^ key
...
\end{lstlisting}%
        \begin{tikzpicture}[remember picture,overlay]
\draw[fill=orange, rounded corners, opacity=0.3]
  ([shift={(-3pt,1.5ex)}]pic cs:seors1) 
  rectangle 
  ([shift={(9pt,-0.65ex)}]pic cs:seors2);

\draw[fill=yellow, rounded corners, opacity=0.3]
  ([shift={(-3pt,1.5ex)}]pic cs:sstr1) 
  rectangle 
  ([shift={(6pt,-0.60ex)}]pic cs:sstr2);

\draw[fill=yellow, rounded corners, opacity=0.3]
  ([shift={(-3pt,1.5ex)}]pic cs:sstr3) 
  rectangle 
  ([shift={(9pt,-0.65ex)}]pic cs:sstr4);
\end{tikzpicture}
        \vspace{-8pt}
\end{tcolorbox}
\end{minipage}
}
      \caption{\label{fig:xor-llvm-sec} Compilation of function \texttt{Xor}
        with no optimizations}
  %% \centering
\end{figure*}

Figure~\ref{fig:xor-sec} shows the mitigation produced by the security
backend of \ac{\toolname} that eliminates leakages that appear in the
LLVM unoptimized code.
The mitigation is based on \textit{instruction scheduling} and
\textit{register allocation} transformations.
In particular, changing the order of operands at line~7 results in a
transition from \texttt{sec} to \texttt{mk} that leaks the value of
\texttt{mask}, which is not secret (marked code at line~7).
Changing the order of the instructions hides the memory-bus leakage.
More specifically, because there are no data dependencies between
lines~3-6, the \texttt{ldr} instruction that causes the leak in
Figure~\ref{fig:xor-llvm} may be scheduled earlier (line~4 in
Figure~\ref{fig:xor-sec}).
Then, another memory instruction that stores the secret value in
memory (line~6 in Figure~\ref{fig:xor-sec}) is scheduled just before
the store instruction at line~8.
This causes a transition from \texttt{sec} to \texttt{mk} in the
memory bus that leaks the value of \texttt{mask} (marked
  code at lines~6 and 8).
These transformations are global, considering possible available
memory instructions and register assignments to mitigate transitional
leakages in the whole program and may (as in Figure~\ref{fig:xor-sec})
introduce no overhead.

However, unoptimized code leads to poor performance.
In general, compiler optimizations may invalidate high-level software
mitigations~\cite{vu_secure_2020}.
Fortunately, this is not the case for register promotion
(\textit{mem2reg} in LLVM), a simple high-level optimization that
enables efficient register allocation by promoting program variables
from memory to registers.
This transformation replaces stack operations to register
operations and preserves the operand order.
In particular, aggressive optimizations (-O1 to -O3 in LLVM) may take
advantage of the associativity property of $\oplus$ to change the
order of the operands, converting $(\texttt{mask} \oplus \texttt{sec})
\oplus \texttt{pub}$ to $\texttt{mask} \oplus (\texttt{sec} \oplus
\texttt{pub})$, which invalidates masking.
%% The register promotion optimization preserves the instruction ordering
%% and does not remove any redundant operations.
%
Equipped with improved high-level code, the \ac{\toolname} backend 
optimizes low-level transformations and generates optimized code.
Figures~\ref{fig:xor-mem2reg-llvm} and \ref{fig:xor-mem2reg-sec} show
the code of Figure~\ref{lst:masked} compiled with register promotion.
Figure~\ref{fig:xor-mem2reg-llvm} leaks the same secret information as
Figure~\ref{fig:xor-llvm} due to register reuse, namely the first
exclusive OR operation \texttt{eors}, but contains no memory-bus
secret leak.
To mitigate the register-reuse leak at line~2, \ac{\toolname} changes
the order of the arguments and the result is now stored in register
\texttt{r2}.

As we see in Figure~\ref{fig:xor-llvm}, unoptimized code may introduce
additional leaks due to the heavy use of the program stack.
Instead, \ac{\toolname} uses register promotion to remove unnecessary memory
accesses that may cause additional leaks.
Then, \ac{\toolname}'s backend generates low-level optimized code that
does not expose secret information through transitional leakages and
does not introduce significant overhead compared to non-secure code.

\section{Threat Model and Modeling Background}
\label{sec:background}

This section describes the \ac{HD} model (Section~\ref{ssec:hdm}), the threat
model (Section~\ref{ssec:threatmodel}), an \ac{HD}-based type-inference
algorithm (Section~\ref{sec:hd-typinf}), a constraint-based compiler backend
model (Section~\ref{ssec:cpbackend}), and the running example for the
constraint-based compiler backend (Section~\ref{ssec:motivation}),

%% \begin{figure}
%%   \input{figs/masked_xor_example}
%%   \caption{\label{lst:masked} Masked exclusive or implementation}
%% \end{figure}

%% \subsection{ARM Cortex M0}

\subsection{Hamming-Distance Model}
\label{ssec:hdm}
The \ac{HW}
model~\cite{messerges1999investigations,kocher_differential_1999,brier_correlation_2004}
corresponds to the number of active bits in a data word.
We assume the following encoding of the binary data, $d =
\sum_{i=0}^{N-1} 2^i d_i$, where $d_i$ is one if the $i_{th}$ bit of an
N-bit word is set and zero otherwise.
The \ac{HW} of this data is the number of bits that are set: $HW(d) =
\sum_{i=0}^{N-1}d_i$.
%
%% \todo{Add some things about the hardware state and the leakage from
%%   the machine.}
%
The \ac{HD} leakage model assumes
that the observed leakage when flipping the bits of a memory element
from a value $d_1$ to a value $d_2$ is  $HW(d_1 \oplus d_2)$,
where $\oplus$ denotes the exclusive OR operation.
If one of the values $d_1$ is a uniform random variable, then $d_1
\oplus d_2$ is also a uniform random variable and $HW(d_1\oplus d_2)$
has the same mean and variance as
$HW(d_1)$~\cite{brier_correlation_2004}.
This means that by masking (exclusive bitwise OR) a secret value $k$
with a uniform random variable $m$, the \ac{HD} of the new
variable has the same mean and variance as $m$.
In this way, masking hides the information of $k$ from the power
consumption traces.

We assume a program $P(\textbf{IN}) = i_1; i_2; ..., i_n$ that takes
as input a set of variables $\textbf{IN}$ and consists of a sequence
of $n$ instructions $i_j$.
We assume that the program has a leakage at every execution step when
there is bit flipping in the hardware registers or the memory bus.
We will use the terms by
\citeauthor{papagiannopoulos_mind_2017}~\cite{papagiannopoulos_mind_2017}
and refer to the hardware-register transition leakage as \ac{ROT} and
the memory-bus transition leakage as \ac{MRE}.
For \ac{MRE}, we assume that both read and write operations make use
of the same memory bus and that the source of the leakage is the
transitional effect when writing the data to the memory bus.
In our model, the memory address of the operations does not affect
the leakage.

\begin{figure*}[!t]
% ensure that we have normalsize text
\normalsize
% Store the current equation number.
%% \setcounter{MYtempeqncnt}{\value{equation}}
%% % Set the equation number to one less than the one
%% % desired for the first equation here.
%% % The value here will have to changed if equations
%% % are added or removed prior to the place these
%% % equations are referenced in the main text.
%% \setcounter{equation}{0}

\begin{flalign}
  &L(P';r \leftarrow e_2; P''; r \leftarrow e_1) &=& L(P';r \leftarrow e_2; P'') \cup \{\mathit{HW}(e_1 \oplus e_2)\}, \nexists i \in P''.~ i = r \leftarrow e_3 \label{eq:leakage1}\\
    &L(P';i_{1};P''; \texttt{mem}(e_b,e_2)) &= & L(P'; i_{1}; P'') \cup \{ \mathit{HW}(e_1 \oplus e_2)\}, (i_{1} = \texttt{mem}(e_a,e_1) ) ~\land~\nexists i \in P''.~ i = \texttt{mem}(e_c, e_3) \label{eq:leakage2}\\
  &L(P';r \leftarrow e) &=&  L(P') \cup \{ \mathit{HW}(e \oplus r_{IN})\},
  \nexists i \in P' . ~i = r \leftarrow e_3 \label{eq:leakage3}\\
  &L(P';\texttt{mem}(e_a,e_1)) &=&  L(P') \cup \{ \mathit{HW}(e_1)\}, 
  \nexists i \in P'.~ i = \texttt{mem}(e_b, e_3)\label{eq:leakage4}
&& 
\end{flalign}

% Restore the current equation number.
%% \setcounter{equation}{\value{MYtempeqncnt}}
% IEEE uses as a separator
\hrulefill
% The spacer can be tweaked to stop underfull vboxes.
\vspace*{4pt}
\end{figure*}

We represent the leakage as a set of observations in the power trace.
To calculate the observed leakage $L(P(\mathit{IN}))$ for an instance
$\mathit{IN}$ of the input variables, we use the \ac{HD} leakage
model.
We write $P = P'; i_n$ to denote a program $P =
i_1;i_2;...;i_{n-1};i_n $, with a prefix $P' = i_1;i_2;...;i_{n-1}$
($\mathit{IN}$ is omitted for simplicity).
%
%% In a similar way, we denote the leakage of a program P as $L(P) =
%% \{L(P'), l_n\}$ a set consisting of $L(P) = \{l_1,...,l_{n},
%% l_{n-1}\}$, with $L(P') = \{l_1,...,l_{n-1}\}$, where $P'$ is a suffix
%% of P and $L(P')$ is the leakage of $P'$.
%
  Equations~\ref{eq:leakage1}-\ref{eq:leakage4} present a recursive
  definition of the leakage model, where for every point in the
  execution trace, the attacker observes the $HW$ of any \ac{ROT} or
  \ac{MRE} transitions.
In the formulas, an expression $e$ is $e \coloneqq r ~|~ v ~|~
\texttt{bop}(e_1,e_2) ~|~ \texttt{uop}(e_1) ~|~ \texttt{mem}(e_a)$,
where $r$ is a register, $v$ is a constant value, \texttt{bop} is a
binary operation, \texttt{uop} is a unary operation, and
$\texttt{mem}(e_a)$ is a memory load operation that loads data from address
$e_a$.
An instruction is $i= r \leftarrow e~|~\texttt{mem}(e_a,e)$, where $r
\leftarrow e$ denotes that an expression is assigned to register $r$,
and $\texttt{mem}(e_a,e)$ is a store memory operation that stores data
$e$ at memory address $e_a$.
To simplify the leakage equations, we transform the load operation
from $r \leftarrow \texttt{mem}(e_a)$ to a sequence
$\texttt{mem}(e_a,v_{\texttt{mem}(e_a)}); r \leftarrow
v_{\texttt{mem}(e_a)}$, where $v_{\texttt{mem}(e_a)}$ is the value in
memory at address $e_a$.
Equation~\ref{eq:leakage1} describes the leakage when two instructions
write the value of their result to the same register and no other
instruction between them writes to the same register.
Note that the first equation deals also with instructions in the form
$r_1 \leftarrow \texttt{bop}(r_2, r_3)$, where \texttt{bop} is a
binary operation and $r_1 = r_2$.
These two-address instructions are common in ARM Thumb and x86
architectures.
Equation~\ref{eq:leakage2} describes the memory-bus leakage of a
memory instruction that writes a value to the memory, given that
another memory instructions precedes this memory instruction.
Equation~\ref{eq:leakage3} describes the leakage of the first instruction
that writes to register $r$.
In this case, the leakage is equal to the \ac{HD} between the new value
and the initial value in register r, $r_{IN}$.
Similarly, Equation~\ref{eq:leakage4} describes the leakage of the
first memory operation.
Here, we assume that the initial memory-bus content, $mb_{IN}$, is a
constant value.
%
%% Finally, Equations~\ref{eq:leakage4}-\ref{eq:leakage6} describe the
%% leakage when no (known) preceding operations write to the same
%% register or use the memory bus.
%
For example, after executing the last instruction of program 
$P = r_1 \leftarrow v_1; \texttt{mem}(v_a, v_2); r_1 \leftarrow v_3; \texttt{mem}(v_b, r_1)$,
the leakage  is equal to
$L(P) \overset{Eq. \ref{eq:leakage2}}{=\joinrel=} L(r_1 \leftarrow v_1; \texttt{mem}(v_a, v_2); r_1 \leftarrow v_3)\cup \{HW( v_3 \oplus v_2)\}
\overset{Eq. \ref{eq:leakage1}}{=\joinrel=} L(r_1 \leftarrow v_1; \texttt{mem}(v_a, v_2))\cup \{HW( v_3 \oplus v_2), HW(v_3 \oplus v_1)\}
\overset{Eq. \ref{eq:leakage4}}{=\joinrel=} L(r_1 \leftarrow v_1)\cup \{HW( v_3 \oplus v_2), HW(v_3 \oplus v_1), HW(v_2)\}
\overset{Eq. \ref{eq:leakage3}}{=\joinrel=} \{HW( v_3 \oplus v_2), HW(v_3 \oplus v_1), HW(v_2), HW(v_1 \oplus r_{1,IN})\}
$, where $r_{1,IN}$ is the initial value of register $r_1$.

Here, we consider that a program is a straight-line function.
Additional checks at the call site are necessary for ensuring the
absence of leakage during function calls, for example to make sure
that the initial memory-bus value is constant.

\subsection{Threat Model}
\label{ssec:threatmodel}
We assume that the software runs on an non-speculative hardware
architecture.
The attacker has access to the software implementation
and the \textit{public} data but not the \textit{secret} data.
The goal of the attacker is to extract information about the secret data
by measuring the power consumption of the device that the code runs on.
The attacker may accumulate a number of traces from multiple runs of
the program and perform statistical analysis, such as
\ac{DPA}~\cite{kocher_differential_1999}.
At every execution, new \textit{random} values are generated and the
attacker has no knowledge of the values of these variables.
Our goal is to eliminate any statistical dependencies between the
secret data and the measured power traces.

We assume that input variables are \texttt{Secret}, \texttt{Public},
or \texttt{Random}.
\texttt{Secret} variables contain sensitive values (e.g.~cryptographic
keys), which the attacker wants to retrieve information about.
\texttt{Public} variables contain values that the attacker knows or
may learn without causing a leakage.
Finally, \texttt{Random} variables follow the uniform distribution in
the domain of the corresponding program variable.
We define the \textit{Leakage Equivalence} security condition for the
generated programs as follows:

\begin{definition}[Leakage Equivalence]
  \label{def:security_condition}
  
Given a program $P(\textbf{IN})$ that has a set of secret input
variables, $\textbf{IN}_{\mathit{sec}} \subseteq \textbf{IN}$, a set of random
input variables, $\textbf{IN}_{\mathit{rand}} \subseteq \textbf{IN}$, and a set
of public input variables, $\textbf{IN}_{\mathit{pub}} \subseteq \textbf{IN}$.
We assume two instances of the input variables, $\mathit{IN}$ and
$\mathit{IN'}$.
%
%% For all instances in each set, we have the same secret and public
%% variables $\forall i, j \in [1,n] ~.~ IN_{sec,i} = IN_{sec,j}~\land~
%% IN_{pub,i} = IN_{pub,j}$.
%
These two instances differ with regards to the set of secret variables
$\mathit{IN}_{\mathit{sec}}$ and $\mathit{IN'}_{\mathit{sec}}$, i.e. for all public variables, $\forall
v\in \mathit{IN}_{\mathit{pub}}$ and $\forall v'\in \mathit{IN'}_{\mathit{pub}}$ we have
$v=v'$.
Let $r \in \mathit{IN}_{\mathit{rand}}$ and $r' \in \mathit{IN'}_{\mathit{rand}}$ be sampled from
a uniform random distribution.
Let $L_p = L(P(\mathit{IN}))$ and $L'_p = L(P(\mathit{IN'}))$.
Then, we say that a program is leakage equivalent if the distributions
of the leakage of the two executions do not
differ, i.e.\ %% $\forall l \in L(P(IN))~ U$
\[\sum_{l \in L_p} \mathbb{E}[l]  = \sum_{l' \in L'_p} \mathbb{E}[l']~\land~
\sum_{l \in L_p} \mathit{Var}(l)  = \sum_{l' \in L'_p} \mathit{Var}(l'),\]
where $\mathbb{E}[l]$ and $\mathit{Var}(l)$ are $l$'s expected value and variance.
\end{definition}

%% For this condition to hold, the leakage of the program should not
%% depend on \texttt{Secret} variables.

\subsection{HD-based Vulnerability Detection}
\label{sec:hd-typinf}
%% \begin{figure*}
%%   \centering
%% \input{figs/llvm_sec_diagram.tex}
%% \caption{\label{fig:wang} \ac{HD} vulnerability detection with
%%   \citeauthor{wang_mitigating_2019-1}~\cite{wang_mitigating_2019-1}}
%% \end{figure*}

In our approach, we need a technique to identify whether two values
result in a \ac{ROT} or and \ac{MRE} leak.
There are different ways to identify whether there is a leak at some
part of the code.
One approach is to use symbolic
execution~\cite{bayrak_sleuth_2013,eldib_formal_2014}.
Symbolic execution executes different paths of a program symbolically
and verifies or invalidates specific properties with the help of
\ac{SMT} solvers.
Symbolic execution is accurate but has scalability issues when the
number of problem variables or program paths increases.
On the other end, type-based
approaches~\cite{gao_verifying_2019-1,wang_mitigating_2019-1} are
typically efficient but at the price of accuracy.
In particular, \citeauthor{wang_mitigating_2019-1} consider a
hierarchy of three types based on the properties of the distribution
they follow: \textit{uniformly random distribution}, \textit{secret
  independent distribution}, or finally \textit{unknown distribution}.
We call these, \texttt{Random}, \texttt{Public}, and \texttt{Secret},
respectively.
The type-inference algorithm assigns a type to each program variable.
To infer the program variable types,
\citeauthor{wang_mitigating_2019-1} define a logic model and solve it
using an \ac{SMT} solver.
The complexity of this approach is low compared to symbolic execution,
at the price of lower accuracy.
%
% revision is only the last part but I marked for rereading
However, the accuracy is sufficient for loop-free, linearized
programs, a format to which many masked and cryptographic
implementations can be transformed~\cite{wang_mitigating_2019-1}.
Because of this, our approach adapts the aforementioned type-inference
analysis, with some accuracy improvements (see supplementary
material~\cite{appendix}).

\subsection{Constraint-based Compiler Backend}
\label{ssec:cpbackend}
%% \todo{cite some more combinatorial backends}.
%% \todo{introduce p??}.
A compiler backend performs three main low-level
transformations to generate low-level code: instruction selection,
instruction scheduling, and register allocation.
A combinatorial compiler backend~\cite{lozano_combinatorial_2019,
  eriksson_integrated_2012,gebotys_efficient_1997} uses combinatorial
solving techniques to optimize software using the aforementioned
transformations.
Different approaches may implement one or more low-level
transformations.
This section focuses on
\ac{CP}~\cite{rossi2006handbook} as a combinatorial solving technique.

\subsubsection{Constraint Model}
\label{ssec:cpmodel}
The constraint-based compiler backend generates a constraint model
that captures the program semantics, the low-level compiler
transformations, and the hardware architecture.
This paper focuses on two compiler transformations, register
allocation and instruction scheduling, that are crucial for our
mitigation.

Compilers typically model the code using an unbounded number of
\textit{virtual} registers until the register allocation stage.
Register allocation assigns each virtual register to a hardware
register, when possible, or a memory slot on the stack
(\textit{spill}), otherwise.
The latter has a negative effect on code efficiency.
Therefore, register allocation transformations attempt to minimize
this effect, while conforming to constraints, such as the number or
hardware registers and the calling conventions.

Instruction scheduling decides on the order of the instructions in a
program.
A valid instruction schedule satisfies the data dependencies among
instructions and the processor resource constraints.

A constraint-based compiler backend may be modeled as a \ac{COP}, $P
=\langle V, U, C, O\rangle$, where $V$ is the set of decision
variables of the problem, $U$ is the domain of these variables, $C$ is
the set of constraints among the variables, and $O$ is the objective
function.
A constraint-based backend aims at minimizing $O$, which typically
models the code's execution time or size.

A program is modeled as a set of basic blocks $B$, pieces of code with
no branches apart from the exit.
Each block contains a number of optional operations, $o \in
\mathit{Operations}$, that may be \textit{active} or not.
%
%However, a preprocessing step may split straight-line code into
%segments to improve the scalability by decomposing the problem.
% (Roberto: too much info, hard to digest at this point)
%
$\mathit{Ins}_o$ denotes the set of hardware instructions that
implement operation $o$.
Each operation includes a number of operands $p \in
\mathit{Operands}$, each of which may be implemented by different,
equally-valued temporaries, $t \in \mathit{Temps}$.
Temporaries are either not live or assigned to a register (hardware
register or the stack).

Figure~\ref{lst:unimasked} shows a simplified version of
the constraint-based compiler backend model for
Figure~\ref{lst:masked}.
Temporaries \texttt{t0}, \texttt{t1}, and \texttt{t2} contain the
input arguments \texttt{pub}, \texttt{mask}, and \texttt{key},
respectively.
Copy operations (\texttt{o2, o3, o4, o6, o8}) enable copying program
values from one register to another (or to the stack) and are
critical for providing flexibility in register allocation.
%% increased
%% number of decisions for register allocation because they allow moving
%% the value of each temporary to another temporary variable.
%
For example, \texttt{o2}, allows the copy of the value \texttt{pub}
from \texttt{t0} to \texttt{t3}.
In the final solution, a copy operation may not be active (shown by
the dash in the set of instructions: \texttt{[ -, copy]}).
The two \texttt{xor} operations (\texttt{o5, o7}) take two operands
each, and each of these operands can in its turn use different but
equally-valued temporary variables, e.g.\ \texttt{t1} and \texttt{t4}.

\begin{figure}
  \begin{tcolorbox}[colback=\codebackcolor, colframe=\codebackcolor,
      top=-5pt, bottom = -5pt,
      right = 0pt, left = 0pt]
\begin{lstlisting}[style = unistyle]
o1: in [t0 $\leftarrow$ pub, t1 $\leftarrow$ mask, t2 $\leftarrow$ key]
o2: t3 $\leftarrow$ [-, copy] t0
o3: t4 $\leftarrow$ [-, copy] t1
o4: t5 $\leftarrow$ [-, copy] t2
o5: t6 $\leftarrow$ xor [t1,t4] [t2,t5]
o6: t7 $\leftarrow$ [-, copy] t6
o7: t8 $\leftarrow$ xor [t0,t3] [t6,t7]
o8: t9 $\leftarrow$ [-, copy] t8
o9: out [t10 $\leftarrow$ [t8,t9]]
\end{lstlisting}
\end{tcolorbox}
\caption{\label{lst:unimasked} Simplified model of the function
  in Figure~\ref{lst:masked}}
\end{figure}

Figure~\ref{lst:unimasked2} shows a valid solution to the register
allocation of the constraint model in Figure~\ref{lst:unimasked}.
All copy operations are deactivated and \texttt{t0}, \texttt{t1}, and
\texttt{t2} are assigned to registers \texttt{R0}, \texttt{R1}, \texttt{R2}.
Temporary \texttt{t6} is assigned to \texttt{R1} and temporary
\texttt{t8} is assigned to \texttt{R0}.
This register assignment is problematic because it induces a
transition in register \texttt{R1} from the initial value that holds
the \texttt{mask} to the masked value \texttt{mask $\oplus$ key},
which leads to a leakage
$L(\texttt{R1} \leftarrow \texttt{R1} \oplus \texttt{R2}; \texttt{R0} \leftarrow \texttt{R0} \oplus \texttt{R1})
\overset{Eq. \ref{eq:leakage3}}{=\joinrel=} L(\texttt{R1}\leftarrow \texttt{R1} \oplus \texttt{R2}) \cup \{HW(\texttt{pub}\oplus (\texttt{pub} \oplus \texttt{mask} \oplus \texttt{key}))\}
\overset{Eq. \ref{eq:leakage3}}{=\joinrel=} \{HW(\texttt{mask}\oplus(\texttt{mask} \oplus \texttt{key})), HW(\texttt{mask} \oplus
\texttt{key})\} = \{HW(\texttt{key}), HW(\texttt{key} \oplus \texttt{mask})\}$.
The first element of the leakage reveals information about \texttt{key}.

The model of instruction scheduling assigns issue cycles to each
operation.
This assignment imposes an ordering of the operation and is
constrained by the program semantics.
For example, in Figure~\ref{lst:unimasked}, scheduling \texttt{o6}
before \texttt{o5} is not allowed because \texttt{o6} depends on
 \texttt{o5} but scheduling \texttt{o4} before \texttt{o3} is
possible.
In Figure~\ref{fig:xor-sec}, the store instruction at line 6 (that
corresponds to line 5 in Figure~\ref{fig:xor-llvm}) is scheduled after
the load instruction at line 4 (line 6 in Figure~\ref{fig:xor-llvm}).
This is allowed because there is no data dependency between these two
instructions.

\begin{figure}
  \begin{tcolorbox}[colback=\codebackcolor, colframe=\codebackcolor,
      top=-5pt, bottom = -5pt,
      right = 0pt, left = 0pt]
\begin{lstlisting}[style = unistyle]
o1: in [t0:R0, t1:R1, t2:R2]
o5: t6:R1 $\leftarrow$ xor t1:R1 t2:R2
o7: t8:R0 $\leftarrow$ xor t0:R0 t6:R1
o9: out [t10:R0]
\end{lstlisting}
\end{tcolorbox}
\caption{\label{lst:unimasked2} Solution of the model in Figure~\ref{lst:unimasked}}
\end{figure}

%% \begin{figure}
%%   \input{figs/masked_xor_example_unison}
%%   \caption{\label{lst:unimasked} Simplified model of masked exclusive or (xor) function
%%     of Figure~\ref{lst:masked}}
%% \end{figure}

%% \todo{example???}

The decision variables of the constraint problem are:
\begin{itemize}
\item $r(t) \in \mathit{Regs}_t,~ t\in \mathit{Temps}$ denotes the hardware register or stack slot assigned to
  temporary $t$;
\item $a(o) \in [\texttt{false}, \texttt{true}],~ o\in \mathit{Operations}$ denotes whether operation $o$
  is active or not;
\item $i(o) \in \mathit{Ins}_o,~ o\in \mathit{Operations}$ is the instruction that
  implements operation $o$;
\item $c(o) \in [0,maxc],~ o\in \mathit{Operations}$ is the cycle at which an
  operation $o$ is scheduled, bounded by $maxc$, a conservative upper bound
  of the execution time;
\item $y(p) \in \mathit{Temps}_p,~ p\in \mathit{Operands}$ is the selected temporary among all possible temporaries for operand $p$.
\end{itemize}

In addition to these, $l(t) \in [\texttt{false}, \texttt{true}],~t \in
\mathit{Temps}$ represents whether a temporary is live or not, $ls(t)
\in [0,maxc],~t \in \mathit{Temps}$ represents the cycle at which $t$
becomes live, and $le(t) \in [0,maxc],~t \in \mathit{Temps}$
represents the last cycle at which $t$ is live.
An important constraint of register allocation is that the register live
ranges of a specific hardware register $r_i$ do not overlap:
\begin{align}
  &\forall t_1, t_2 \in \mathit{Temps}~.~l(t_1)~\land~l(t_2)~\land~r(t_1) = r(t_2) \implies \nonumber\\
  &~~~~~~~~~~~~~~~~~~~~~~ls(t_1) \ge le(t_2) ~\lor~ ls(t_2) \ge le(t_1).\label{eq:nooverlap}
\end{align}

Moreover, when a temporary is live, its last live cycle ($le$) is
strictly greater than its live start ($ls$):
\begin{align}
  \forall t\in \mathit{Temps}~.~l(t) \implies ls(t) < le(t).
\label{eq:liveend}
\end{align}

\subsubsection{Objective Function}

A typical objective function of a constraint-based backend minimizes
different metrics such as \textit{code size} and \textit{execution time}.
These can be captured in a generic objective function that sums up
the weighted cost of each basic block:
\[
\sum_{b\in B} \mathit{weight}(b) \cdot \textbf{cost}(b).
\]
The \textbf{cost} of each basic block consists of the cost of the specific
implementation and is a variable, whereas \textit{weight} is a
constant value that represents the contribution of the specific basic
block to the total cost.
This cost model is accurate for simple hardware architectures.
However, in the presence of advance microarchitectural features, such
as complex cache hierarchy, branch prediction, and/or out-of-order
execution, the cost model is not accurate.

%% In our example at Figure~\ref{lst:unimasked}, a solution to
%% the problem is 

\subsection{Example in a Constraint-based Compiler Backend}
\label{ssec:motivation}
\begin{figure*}
  \centering
  \input{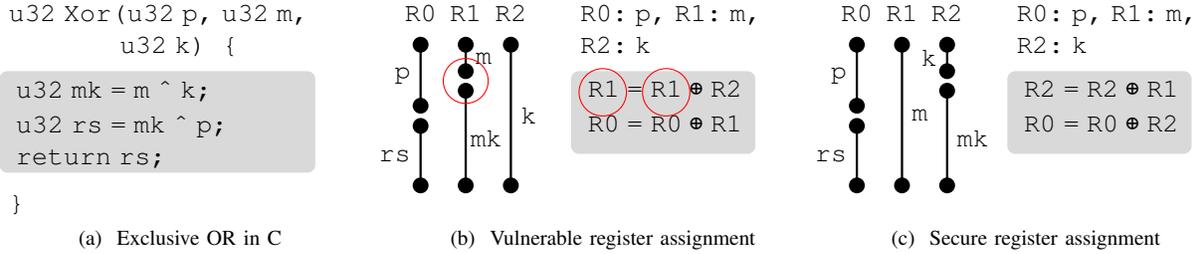}%
\newcommand{\rzero}{\texttt{R0}}
\newcommand{\rone}{\texttt{R1}}
\newcommand{\rtwo}{\texttt{R2}}
\newcommand{\rthree}{\texttt{R3}}
\newcommand{\cs}{\hspace{0.1cm}}
\begin{SaveVerbatim}[commandchars=\\\[\]]{fac-in}
u32\cs[]Xor(u32\cs[]p,\cs[]u32\cs[]m,
       u32\cs[]k) {
\end{SaveVerbatim}
%% \begin{SaveVerbatim}[commandchars=\\\[\]]{fac-b1}
%% \cs[]u32\cs[]mk\cs[]=\cs[]m\cs[]\ttoplus\cs[]k;
%% \end{SaveVerbatim}
\begin{SaveVerbatim}[commandchars=\\\[\]]{fac-b1}
\cs[]u32\cs[]mk\cs[]=\cs[]m\cs[]^\cs[]k;
\end{SaveVerbatim}
\begin{SaveVerbatim}[commandchars=\\\[\]]{fac-b2s1}
\cs[]u32\cs[]rs\cs[]=\cs[]mk\cs[]^\cs[]p;
\end{SaveVerbatim}
\begin{SaveVerbatim}[commandchars=\\\[\]]{fac-b2s2}
\cs[]return\cs[]rs;
\end{SaveVerbatim}
%% \begin{SaveVerbatim}[commandchars=\\\[\]]{fac-b2s3}
%% \cs[]\cs[]n--;
%% \end{SaveVerbatim}
%% \begin{SaveVerbatim}[commandchars=\\\[\]]{fac-b2s4}
%% \cs[]}
%% \end{SaveVerbatim}
\begin{SaveVerbatim}{fac-out}
}
\end{SaveVerbatim}

\begin{SaveVerbatim}[commandchars=\\\[\]]{ass1-in}
R0:\cs[]p,\cs[]R1:\cs[]m,
R2:\cs[]k
\end{SaveVerbatim}
\begin{SaveVerbatim}[commandchars=\\\[\]]{ass1-b1}
\cs[]R1\cs[]=\cs[]R1\cs[]\ttoplus\cs[]R2
\end{SaveVerbatim}
\begin{SaveVerbatim}[commandchars=\\\[\]]{ass1-b2s1}
\cs[]R0\cs[]=\cs[]R0\cs[]\ttoplus\cs[]R1
\end{SaveVerbatim}
%% \begin{SaveVerbatim}[commandchars=\\\[\]]{ass1-b2s2}
%% \cs[]return\cs[]rs;
%% \end{SaveVerbatim}
\begin{SaveVerbatim}{ass1-out}
OUT: [R0]
\end{SaveVerbatim}

\begin{SaveVerbatim}[commandchars=\\\[\]]{ass2-in}
R0:\cs[]p,\cs[]R1:\cs[]m,
R2:\cs[]k
\end{SaveVerbatim}
\begin{SaveVerbatim}[commandchars=\\\[\]]{ass2-b1}
\cs[]R2\cs[]=\cs[]R2\cs[]\ttoplus\cs[]R1
\end{SaveVerbatim}
\begin{SaveVerbatim}[commandchars=\\\[\]]{ass2-b2s1}
\cs[]R0\cs[]=\cs[]R0\cs[]\ttoplus\cs[]R2
\end{SaveVerbatim}
\begin{SaveVerbatim}{ass2-out}
OUT: [R0]
\end{SaveVerbatim}

\newcommand{\ws}{\hspace{0.1cm}}
\newcommand{\asmlabel}[1]{{\it{#1}}}
\newcommand{\asmbundled}{$\hspace{0.17cm}\bundled{}\hspace{0.12cm}$}

\begin{tikzpicture}
  % C code
  \coordinate (bin);
  \coordinate [below=1.0cm of bin]  (b1s1);
  \coordinate [below=0.46cm of b1s1] (b1s2);
  \coordinate [below=0.46cm of b1s2] (b1s3);
  \coordinate [below=0.6cm of b1s3]   (bout);
  
  \newcommand{\codeBlockWidth}{4.2cm}  

  \node [cfg block, right=of bin, minimum height=1.0cm, invisible, anchor = north west] (blockin) {};
  \node [below right] at (blockin.north west) (v0) {\BUseVerbatim{fac-in}};

  \node [cfg block,right=of b1s1, minimum width=\codeBlockWidth, minimum height=1.35cm, anchor = north west] (block1s1) {};
  \node [below right] at (block1s1.north west) (v1) {\BUseVerbatim{fac-b1}};
  \node [cfg block, right=of b1s2, minimum height=0.5cm, invisible, anchor = north west] (block1s2) {};
  \node [below right] at (block1s2.north west) (v2) {\BUseVerbatim{fac-b2s1}};
  \node [cfg block, right=of b1s3, minimum height=0.5cm, invisible, anchor = north west] (block1s3) {};
  \node [below right] at (block1s3.north west) (v3) {\BUseVerbatim{fac-b2s2}};
  \node [cfg block, right=of bout, minimum height=0.5cm, invisible, anchor = north west] (blockout) {};
  \node [below right] at (blockout.north west) (v4) {\BUseVerbatim{fac-out}};
  % Program points
  \coordinate (entry)  at (blockin.north west);
  \coordinate (bin-p1) at (blockin.south west);
  % Between program points
  \coordinate [below=0.1cm of bin-p1] (bin-entry);

% registers
  \newcommand{\vulnRZeroDistance}{5.6cm}
  \newcommand{\registerDistance}{0.6cm}
  \coordinate [right=\vulnRZeroDistance of entry] (vulnRZero);
  \node [below=0cm of vulnRZero] {\texttt{\rzero{}}};
  \coordinate [right=\registerDistance of vulnRZero] (vulnROne);
  \node [below=0cm of vulnROne] {\texttt{\rone{}}};
  \coordinate [right=\registerDistance of vulnROne] (vulnRTwo);
  \node [below=0cm of vulnRTwo] {\texttt{\rtwo{}}};

  % p
  \coordinate (vulnr00) at (vulnRZero |- v0.-5);
  \coordinate (vulnr01) at (vulnRZero |- v2.north);

  \begin{scope}[live range]
    \draw [<->] (vulnr00) -- node [left] {\texttt{p}}(vulnr01);
  \end{scope}
  % rs
  \coordinate (vulnr00-rs) at (vulnRZero |- v2.center);
  \coordinate (vulnr01-rs) at (vulnRZero |- v4.north);

  \begin{scope}[live range]
    \draw [<->] (vulnr00-rs) -- node [left] {\texttt{rs}}(vulnr01-rs);
  \end{scope}
  % k
  \coordinate (vulnr10-k) at (vulnRTwo |- v0.-5);
  \coordinate (vulnr11-k) at (vulnRTwo |- v4.north);

  \begin{scope}[live range]
    \draw [<->] (vulnr10-k) -- node [right] {\texttt{k}} (vulnr11-k);
  \end{scope}
  % m
  \coordinate (vulnr20-m) at (vulnROne |- v0.-5);
  \coordinate (vulnr21-m) at (vulnROne |- v1.north);

  \begin{scope}[live range]
    \draw [<->] (vulnr20-m) -- node [right] {\texttt{m}} (vulnr21-m);
  \end{scope}
  % mk
  \coordinate (vulnr20-mk) at (vulnROne |- v1.center);
  \coordinate (vulnr21-mk) at (vulnROne |- v4.north);

  \begin{scope}[live range]
    \draw [<->] (vulnr20-mk) --node [right=-2pt] {\texttt{mk}} (vulnr21-mk);
  \end{scope}

  % leak

  \node[circle,draw,red, text width = 9pt] (c) at
                         ($(vulnROne |- v1.north)!0.5!(vulnROne |- v1.center)$){};

  % Register assignment vuln
  \coordinate [right=7.6cm of bin] (bin1);
  \coordinate [below=1.0cm of bin1]  (b1s11);
  \coordinate [below=0.46cm of b1s11] (b1s21);
  \coordinate [below=0.46cm of b1s21] (b1s31);
  \coordinate [below=0.6cm of b1s31]   (bout1);
  
  \newcommand{\codeBlockWidthone}{2.45cm}  

  \node [cfg block, right=of bin1, minimum height=1.0cm, invisible, anchor = north west] (blockin1) {};
  \node [below right] at (blockin1.north west) (v01) {\BUseVerbatim{ass1-in}};
  \node [cfg block,right=of b1s11, minimum width=\codeBlockWidthone, minimum height=1.1cm, anchor = north west] (block1s11) {};
  \node [below right] at (block1s11.north west) (v11) {\BUseVerbatim{ass1-b1}};
  \node [cfg block, right=of b1s21, minimum height=0.5cm, invisible, anchor = north west] (block1s21) {};
  \node [below right] at (block1s21.north west) (v21) {\BUseVerbatim{ass1-b2s1}};
  %% \node [cfg block, right=of bout, minimum height=0.5cm, invisible, anchor = north west] (blockout) {};
  %% \node [below right] at (blockout.north west) (v41) {\BUseVerbatim{ass1-out}};

  \coordinate (r2) at ([xshift=12pt,yshift=-8pt]v11.north west);
  \node[circle, draw, red, text width = 10pt] (c) at (r2) {};
  \node[circle, draw, red, text width = 10pt] (c) at ([xshift=24pt]r2) {};

  % Non vulnerable
  \newcommand{\secRZeroDistance}{11.4cm}
  \coordinate [right=\secRZeroDistance of entry] (secRZero);
  \node [below=0cm of secRZero] {\rzero{}};
  \coordinate [right=\registerDistance of secRZero] (secROne);
  \node [below=0cm of secROne] {\texttt{\rone{}}};
  \coordinate [right=\registerDistance of secROne] (secRTwo);
  \node [below=0cm of secRTwo] {\texttt{\rtwo{}}};
  % p
  \coordinate (secr00) at (secRZero |- v0.-5);
  \coordinate (secr01) at (secRZero |- v2.north);

  \begin{scope}[live range]
    \draw [<->] (secr00) -- node [left] {\texttt{p}}(secr01);
  \end{scope}
  % rs
  \coordinate (secr00-rs) at (secRZero |- v2.center);
  \coordinate (secr01-rs) at (secRZero |- v4.north);

  \begin{scope}[live range]
    \draw [<->] (secr00-rs) -- node [left] {\texttt{rs}}(secr01-rs);
  \end{scope}
  % k
  \coordinate (secr10-k) at (secRTwo |- v0.-5);
  \coordinate (secr11-k) at (secRTwo |- v1.north);

  \begin{scope}[live range]
    \draw [<->] (secr10-k) -- node [left] {\texttt{k}} (secr11-k);
  \end{scope}
  % m
  \coordinate (secr20-m) at (secROne |- v0.-5);
  \coordinate (secr21-m) at (secROne |- v4.north);

  \begin{scope}[live range]
    \draw [<->] (secr20-m) -- node [right] {\texttt{m}} (secr21-m);
  \end{scope}
  % mk
  \coordinate (secr20-mk) at (secRTwo |- v1.center);
  \coordinate (secr21-mk) at (secRTwo |- v4.north);

  \begin{scope}[live range]
    \draw [<->] (secr20-mk) -- node [right] {\texttt{mk}} (secr21-mk);
  \end{scope}

  % Register assignment sec
  \coordinate [right=13.4cm of bin] (bin2);
  \coordinate [below=1.0cm of bin2]  (b1s12);
  \coordinate [below=0.46cm of b1s12] (b1s22);
  \coordinate [below=0.46cm of b1s22] (b1s32);
  \coordinate [below=0.6cm of b1s32]   (bout2);
  
  \newcommand{\codeBlockWidthtwo}{2.45cm}
  \node [cfg block, right=of bin2, minimum height=1.0cm, invisible, anchor = north west] (blockin2) {};
  \node [below right] at (blockin2.north west) (v02) {\BUseVerbatim{ass2-in}};
  \node [cfg block,right=of b1s12, minimum width=\codeBlockWidthtwo, minimum height=1.1cm,
  anchor = north west] (block1s12) {};
  \node [below right] at (block1s12.north west) (v12) {\BUseVerbatim{ass2-b1}};
  \node [cfg block, right=of b1s22, minimum height=0.5cm, invisible, anchor = north west] (block1s22) {};
  \node [below right] at (block1s22.north west) (v22) {\BUseVerbatim{ass2-b2s1}};

\end{tikzpicture}
  \subfloat[\label{fig:xor-c} Exclusive OR in C]{\hspace{.31\linewidth}}
  \subfloat[\label{fig:xor-ra-vuln} Vulnerable register assignment]{\hspace{.31\linewidth}}
  \subfloat[\label{fig:xor-ra-sec} Secure register assignment]{\hspace{.31\linewidth}}
  %% \subfloat[\label{fig:xor-ra-diff} Alternative secure register assignment]{\hspace{.2\linewidth}}
  \caption{\label{fig:xor-motivation} The exclusive OR example, illustrating a 
\ac{HD} vulnerability and alternative register assignments
%
%% \roberto{Nice figure! I think it would help if the register-assigned code for
%%   each case was also shown, to make the correspondence clearer.}
  }
\end{figure*}

%% Register allocation and instructions scheduling affect the
%% security of a masked program by enabling\ac{HD} vulnerabilities

%% \roberto{I think this would fit better in the introduction, see my
%%   comments there.}
%
Low-level transformations, like register allocation and instruction
scheduling, affect the security of programs.
Figure~\ref{fig:xor-c} shows the high-level masked implementation of
exclusive OR in C (same as Figure~\ref{lst:masked}).
The code takes three inputs: \texttt{p} (a \texttt{Public} value),
\texttt{k} (a \texttt{Secret} value), and \texttt{m} (a
\texttt{Random} variable).
The code computes first the exclusive OR of \texttt{m} and \texttt{k}
and stores it in \texttt{mk}.
Then, it computes the exclusive OR of \texttt{mk} with \texttt{p} and
stores it in \texttt{rs}, which the function returns.

Figure~\ref{fig:xor-ra-vuln} shows a register allocation of function
\texttt{Xor} that leads to a \ac{HD} vulnerability.
Both \texttt{m} and \texttt{mk} are stored in the same register,
hence the content of \texttt{mk} replaces the previous value
\texttt{m} in register \texttt{R1}.
According to the leakage model, the attacker observes the exclusive
OR between the initial and updated value of a hardware register.
Using the register allocation of Figure~\ref{fig:xor-ra-vuln}, the
leakage reveals information about the secret: $\ac{HW}(\texttt{mk}
\oplus \texttt{m}) = \ac{HW}((\texttt{m} \oplus \texttt{k}) \oplus
\texttt{m}) = \ac{HW}(\texttt{k})$.
Value \texttt{k} is a secret value, and thus, a leak occurs (circled in
Figure~\ref{fig:xor-ra-vuln}).

A constraint-based compiler backend is able to generate all legal
register allocations for a program.
Figure~\ref{fig:xor-ra-sec} shows an alternative register allocation
for function \texttt{Xor}.
Here, the result of \texttt{mk} is written in hardware register
\texttt{R2}, giving a \ac{HD} leakage $\ac{HW}(\texttt{mk} \oplus
\texttt{k}) = \ac{HW}((\texttt{m} \oplus \texttt{k}) \oplus
\texttt{k}) = \ac{HW}(\texttt{m})$.
The leakage here corresponds to the value of \texttt{m}, which is not
a sensitive value.
%
%% Different secure solutions are also possible.
%
%% For example, Figure~\ref{fig:xor-ra-diff} shows an alternative
%% register allocation using a \texttt{copy} operation for function
%% \texttt{Xor}.
%% %
%% Here, \texttt{k} is copied to register \texttt{R3} and variable
%% \texttt{mk} is assigned to the same register leading to an \ac{HD}
%% leakage equal with $\texttt{mk} \oplus \texttt{k} = (\texttt{m} \oplus
%% \texttt{k}) \oplus \texttt{k} = \texttt{m}$.
%% %
In a similar way, instruction scheduling may be able to remove
leakages as seen in Figure~\ref{fig:xor-llvm-sec}.
By changing the schedule of the instructions, the model is often able
to generate a \ac{PSC}-free solution with no code quality overhead.

This example shows that low-level transformations can be responsible for the
introduction of \ac{HD} vulnerabilities and have thus to be taken into account
to provide effective mitigations.

%% and different valuations
%% $k\subseteq v$ and $k'\subseteq v'$ of these variables, the leakage of the two executions should
%% not differ, i.e.\ $L(P(v)) = L(P(v'))$.

\section{\toolname}

\begin{figure*} %[ht!]
  \centering
  \input{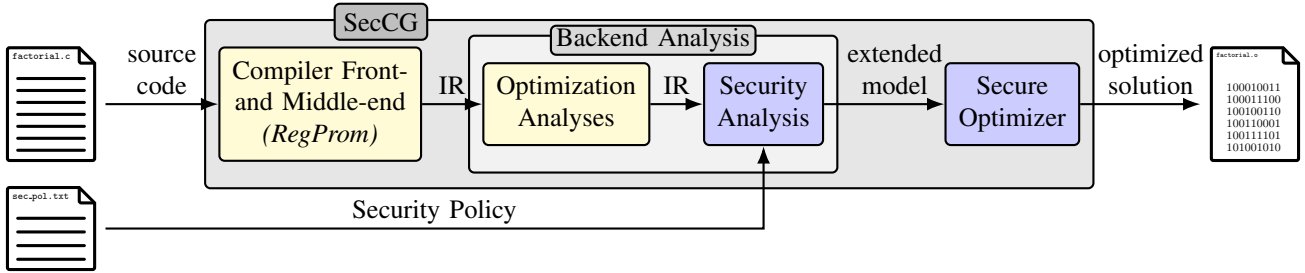}
  \vspace{-5pt}
\caption{\label{fig:comp} High-level view of \ac{\toolname}}
\end{figure*}

This section introduces \ac{\toolname}, an approach to optimize code
that is secure against \ac{PSC} attacks.
%
%% Compiler code generation may generate code that is vulnerable to power
%% side-channel attacks.
%% %
%% To remedy this problem, we propose \ac{\toolname}, an approach to
%% optimize code that does not expose secret information.
%
Figure~\ref{fig:comp} shows the high-level view of \ac{\toolname}.
\ac{\toolname} is a constraint-based optimizing secure compiler,
i.e.\ it extends a constraint-based compiler backend with security
constraints.
It takes two inputs: 1) a C or C++ program, and 2) a
security policy denoting which variables are \texttt{Secret},
\texttt{Random}, or \texttt{Public}.
\ac{\toolname} enables \textit{register promotion} at the compiler
middle end because this optimization preserves the high-level
properties of the program and, at the same time, creates substantial
opportunities for register allocation.
Then, the constraint-based compiler backend, extended with security
constraints, takes as input the program in a machine-level \ac{IR} and
the security policy.
Next, \ac{\toolname} performs a security analysis (see
Section~\ref{sec:hd-typinf}).
The results are used to impose constraints that prevent
\ac{HD} vulnerabilities.
Given the secure model, the approach generates an optimized solution.

Section~\ref{sec:tinf} presents the security analysis.
Section~\ref{sec:constraint_model} presents the secure constraint
model that extends the constraint-based compiler backend.
Finally, Section~\ref{sec:search} presents the solving enhancements of
\ac{\toolname}.

\subsection{Security Analysis}
\label{sec:tinf}

\ac{\toolname} performs a security analysis to extract the security
types of each program variable and, subsequently, generates constraints
that prohibit insecure low-level implementations.
The security analysis identifies the security type (\texttt{Random},
\texttt{Public}, or \texttt{Secret}) of each intermediate variable.
In the compiler constraint model, the program variables correspond to
the input arguments, the operands and the result of each operation.
This is equivalent to the temporary variables, i.e. the virtual
registers.
Each operand can use a number of alternative temporary values $t \in
\mathit{Temps}$ and each temporary value is assigned to a register
(see Section~\ref{ssec:cpbackend}).
The type-inference rules do not handle loops or conditional
statements.
However, cryptographic implementations that are free from \acp{PSC}
are often linearizable~\cite{wang_mitigating_2019-1}.
%
%% Therefore, without loss of generality, we assume that the input program is
%% linear, i.e.\ free from loops and branches.
%% %

The security analysis uses a type-inference algorithm based on
\citeauthor{wang_mitigating_2019-1}~\cite{wang_mitigating_2019-1}.
We extend this algorithm with additional definitions that improve the
accuracy of the type inference (see
  supplementary material~\cite{appendix}).
In particular, we extend the type-inference algorithm with rules that
consider additional properties of GF($2^n$), like distributivity
between exclusive or ($\oplus$) and multiplication in GF($2^n$)
($\odot$).
At the end of the analysis, all temporary variables have an inferred
type.
Figure~\ref{lst:unimasked3} shows the inferred security types for each
of the temporaries in our running example.
Temporaries \texttt{t0} and \texttt{t3} are \texttt{Public} (green),
\texttt{t2} and \texttt{t5} are \texttt{Secret} (red), and \texttt{t1},
\texttt{t4} and \texttt{t6-t10} are \texttt{Random} (brown).

\begin{figure}
  \begin{tcolorbox}[colback=\codebackcolor, colframe=\codebackcolor,
      top=-5pt, bottom = -5pt,
      right = 0pt, left = 0pt]
\begin{lstlisting}[style = typedunistyle]
o1: in [t0:Public, t1:Random, t2:Secret]
o2: t3:Public $\leftarrow$ [-, copy] t0
o3: t4:Random $\leftarrow$ [-, copy] t1
o4: t5:Secret $\leftarrow$ [-, copy] t2
o5: t6:Random $\leftarrow$ xor [t1,t4] [t2,t5]
o6: t7:Random $\leftarrow$ [-, copy] t6
o7: t8:Random $\leftarrow$ xor [t0,t3] [t6,t7]
o8: t9:Random $\leftarrow$ [-, copy] t8
o9: out [t10:Random $\leftarrow$ [t8,t9]]
\end{lstlisting}
\end{tcolorbox}
\caption{\label{lst:unimasked3} Typed intermediate representation}
\end{figure}

The type-inference algorithm is conservative.
Function $\mathit{type}(t): \mathit{Temps} \to \{R, S, P\}$ returns the type assigned to temporary variable $t$.
This section abbreviates the types as
follows: type $R$ corresponds to \texttt{Random}, $S$ corresponds to
\texttt{Secret}, and $P$ corresponds to \texttt{Public}.

In the following, we define the data that the security analysis
provides to the constraint model, which the latter requires to impose
security constraints.
%
%% At the end of this section, we provide a proof sketch that our
%% Given the type inference, the security analysis generates information
%% that the solver uses to generate secure programs.
%
According to the leakage model, when a hardware register
changes from one value to another, the exclusive OR of the two values
is exposed.
$Rpairs$ is the set of temporary variable pairs that when xor:ed
together reveal secret information:
\begin{align}
  Rpairs = \{ &(t_1, t_2) ~|~  t_1 \in \mathit{Temps} \andcons t_2 \in \mathit{Temps} \andcons \nonumber\\
  & (\mathit{type}(t_1) \in \{R,P\}) \andcons (\mathit{type}(t_2) \in \{R,P\}) \andcons \nonumber\\
    &(\mathit{type}(t_1 \oplus t_2) = S) \}.\label{eq:rpairs}
\end{align}
%
%% These pairs of temporary variables, $Rpairs$, should not lead to
%% register-reuse transitional effects, i.e.\ they should not be assigned
%% to the same register one following the other.
%
%
In the running example (Figure~\ref{lst:unimasked3}), $Rpairs = \{
  \allowbreak (\texttt{t1}, \texttt{t6}),
  \allowbreak (\texttt{t1}, \texttt{t7}),
  \allowbreak (\texttt{t1}, \texttt{t8}),
  \allowbreak (\texttt{t1}, \texttt{t9}),
  \allowbreak (\texttt{t4}, \texttt{t6}),
  \allowbreak (\texttt{t4}, \texttt{t7}),
  \allowbreak (\texttt{t4}, \texttt{t8}),
  \allowbreak (\texttt{t4}, \texttt{t9}),
  \allowbreak (\texttt{t6}, \texttt{t7}),
  \allowbreak (\texttt{t6}, \texttt{t8}),
  \allowbreak (\texttt{t6}, \texttt{t9}),
  \allowbreak (\texttt{t7}, \texttt{t8}),
  \allowbreak (\texttt{t7}, \texttt{t9}),
  \allowbreak (\texttt{t8}, \texttt{t9})\} $.
For every pair of temporaries in $Rpairs$, a constraint prohibits the
contiguous assignment of the temporaries to the same register 
(\texttt{m} and \texttt{mk} in Figure~\ref{fig:xor-ra-vuln}).

$Rpairs$ do not consider secret values.
Instead, if the type of a temporary variable $t$ is \texttt{Secret},
we impose a different constraint because the secret information will
always result in a leak.
In this case, we impose the constraint that another random variable
should precede and follow the definition of the secret variable to
mask the secret information.
$Spairs$ is a set of pairs, each of which consists of a secret
temporary variable $t$ and a set of random temporary variables $ts$
that are able to hide the secret information, i.e.\ $\forall t' \in ts
~.~type(t' \oplus t) = R$:
\begin{flalign}
 &Spairs = \{  (t, ts) ~|~ t \in \mathit{Temps} \andcons \mathit{type}(t) = S \andcons\nonumber\\
   &~~~~~~~~ ts = \{ t' ~|~ t' \in \mathit{Temps} \andcons \nonumber\\
   &~~~~~~~~ ~~~~~~~\mathit{type}(t') = R \andcons  \mathit{type}(t' \oplus t) = R\}\}.\label{eq:spairs}&&
\end{flalign}
In the example (Figure~\ref{lst:unimasked3}),
$Spairs = \{(\texttt{t5}, \allowbreak \{\texttt{t4},
    \allowbreak \texttt{t6},
    \allowbreak \texttt{t7},
    \allowbreak \texttt{t8},
    \allowbreak \texttt{t9}\})\}$.

%% \begin{theorem}[Secret Registers]
%%   If $type(t) = S,~ t\in \mathit{Temps} \Longleftrightarrow \exists (t_i,t_s)\in Spairs.~
%%   t_i = t$. \label{th:secregs}
%% \end{theorem}
%% \begin{proof} From the definition of $Spairs$ (Equation~\ref{eq:spairs}).
%% \end{proof}
Memory operations may also reveal secret information.
We assume the same leakage model (\ac{HD} model) for the memory bus as
for the register-reuse transitional effects.
This means that the leakage corresponds to the exclusive OR of two
subsequent memory operations.
$Mmpairs$ includes the pairs of memory operations that result in
memory-bus transitional leakage, i.e. pairs of memory operations that
when scheduled subsequently lead to a secret leakage.
\begin{align}
  Mmpairs = \{ & (o_1, o_2) ~|~ o_1 \in \mathit{MemOperations} \andcons\nonumber \\
    & o_2 \in \mathit{MemOperations} \andcons\nonumber\\
    & \mathit{type}(tm(o_1)) \in \{R,P\} \andcons \nonumber\\
    & \mathit{type}(tm(o_2)) \in \{R,P\} \andcons \nonumber\\
    & \mathit{type}(tm(o_1) \oplus tm(o_2) = S)\}.\label{eq:mmpairs}
\end{align}
Here, $tm(o)\in \mathit{Temps}$ is the temporary that corresponds to the memory
data of the operation.
In the example (Figure~\ref{lst:unimasked3}), $Mmpairs = \{
\allowbreak (\texttt{o3},\texttt{o6}),
\allowbreak (\texttt{o3},\texttt{o8}),
\allowbreak (\texttt{o6},\texttt{o8})\}$, in case $\texttt{o3}$, $\texttt{o6}$,
$\texttt{o8}$, are memory spills.
Note that, for simplicity, Figure~\ref{lst:unimasked3} does not
include all copies for memory spilling as we would need to duplicate
the copies for first storing and then loading the variables.

The same leakage as in the case when a secret value was written to a
register applies here.
If a memory operation stores/loads a secret value to/from the memory,
a random memory operation that is able to hide the secret information
should precede and follow this operation.
$Mspairs$ is a set of pairs, each of which consists of the memory
operation that accesses secret data, $o$, and a set of memory
operations that access random data and are able to hide the secret
information, i.e.\ $\mathit{type}(tm(o') \oplus tm(o)) = R$:
\begin{align}
  Mspairs = \{ & (o, os) ~|~ o \in \mathit{MemOperations} \andcons \nonumber\\
   & \mathit{type}(tm(o)) = S \andcons \nonumber\\
   & os = \{ o' ~|~ o' \in \mathit{MemOperations} \andcons \nonumber\\
   & ~~~~~~~~\mathit{type}(tm(o')) = R \andcons  \nonumber\\
   & ~~~~~~~~\mathit{type}(tm(o') \oplus tm(o)) = R\}\}.\label{eq:mspairs}
\end{align}
In the example (Figure~\ref{lst:unimasked3}), $Mspairs = \{(
  \allowbreak \texttt{o4},
  \allowbreak \{\texttt{o3},
  \allowbreak \texttt{o6},
  \allowbreak \texttt{o8}\})\}$, in case $\texttt{o4}$,
  $\texttt{o3}$, $\texttt{o6}$, and $\texttt{o8}$ are spilled in memory.

The security analysis provides $Rpairs$, $Spairs$, $Mmpairs$, and $Mspairs$ to
the constraint model, which enables constraining code generation to generate
secure implementations.

\subsection{Constraint Model}
\label{sec:constraint_model}

The constraint model takes as input the four sets computed by the security
analysis ($Rpairs$, $Spairs$, $Mmpairs$, and $Mspairs$) and uses them to
generate appropriate constraints that prohibit insecure solutions.
Predicate \texttt{samereg} tells whether the two input temporaries are
active (\texttt{l(t)} = 1) and are assigned to the same register.

% TODO(Romy) add $\land$ is_hr(r(t$_1$)) ?
\begin{tcolorbox}[colback=\colbackcolor,colframe=\colframecolor, top=-5pt, bottom = -5pt,
  right = 0pt, left = 0pt]
\begin{lstlisting}[style = modelingstyle]
pred samereg(t$_1$,t$_2$):
  l(t$_1$) $\land$ l(t$_2$) $\land$ (r(t$_1$) $=$ r(t$_2$)) 
\end{lstlisting}
\end{tcolorbox}

In Figure~\ref{lst:unimasked2}, \texttt{samereg(t0,t8)} \texttt{=}
\texttt{l(t0)} $\land$ \texttt{l(t8)} $\land$ (\texttt{r(t0)} $=$
\texttt{r(t8)}) \texttt{=} \texttt{true}, \texttt{samereg(t2,t6)}
\texttt{=} \texttt{false} (\texttt{r(t2)} $\neq$
\texttt{r(t6)}), and \texttt{samereg(t1,t7)} \texttt{=}
\texttt{false} (\texttt{t7} is not live).

\subsubsection{Rpairs Constraints}
\label{ssec:rconstraint}
The following constraint ensures that a pair of temporaries in
$Rpairs$ are either not assigned to the same register or they are not
subsequent (\texttt{subseq} constraint, defined in
Section~\ref{sec:subseq}).

\begin{tcolorbox}[colback=\colbackcolor,colframe=\colframecolor, top=-5pt, bottom = -5pt,
  right = 0pt, left = 0pt]
\begin{lstlisting}[style = modelingstyle]
forall (t$_1$,t$_2$) in $Rpairs$:
  samereg(t$_1$, t$_2$) $\implies$
    ($\neg$subseq(t$_1$,t$_2$) $\land$ $\neg$subseq(t$_2$,t$_1$))
\end{lstlisting}
\end{tcolorbox}

In Figure~\ref{lst:unimasked2}, this constraint is not satisfied for
\texttt{t1} and \texttt{t6} because \texttt{samereg(t2,t6)} \texttt{=}
\texttt{true} and \texttt{subseq(t2,t6)} \texttt{=} \texttt{true}.

\subsubsection{Spairs Constraints}
\label{ssec:sconstraint}
The following constraint ensures that for each pair
$(\texttt{t}_s,\texttt{t}_{rs}) \in Spairs$, if $\texttt{t}_s$ is
live, one of the random temporaries $\texttt{t}_r \in
{\texttt{t}_{rs}}$ precedes the secret temporary $\texttt{t}_s$ and
another random temporary succeeds $\texttt{t}_s$.

\begin{tcolorbox}[colback=\colbackcolor,colframe=\colframecolor, top=-5pt, bottom = -5pt,
  right = 0pt, left = 0pt]
\begin{lstlisting}[style = modelingstyle]
forall (t$_s$,t$_{rs}$) in $Spairs$:
  exists t$_r$ in t$_{rs}$:
      l(t$_s$) $\implies$ (l(t$_r$) $\land$ subseq(t$_r$,t$_s$))
  $\land$
  exists t$_r$ in t$_{rs}$:
      l(t$_s$) $\implies$ (l(t$_r$) $\land$ subseq(t$_s$,t$_r$))
\end{lstlisting}
\end{tcolorbox}

Figure~\ref{lst:unimasked4} shows a solution to the model in
Figure~\ref{lst:unimasked}, where both the $Rpairs$ and the $Spairs$
constraints are satisfied.
\texttt{t5} is active but is assigned to the same register as
\texttt{t4}, which precedes \texttt{t5} and thus eliminates the
leakage.
Similarly, \texttt{t6} follows the assignment of \texttt{t5} and thus
hides the secret value.

\begin{figure}
  \begin{tcolorbox}[colback=\codebackcolor, colframe=\codebackcolor,
      top=-5pt, bottom = -5pt,
      right = 0pt, left = 0pt]
\begin{lstlisting}[style = unistyle]
o1: in [t0:R0, t1:R1, t2:R2]
o3: t4:R3 $\leftarrow$ t1:R1
o4: t5:R3 $\leftarrow$ t2:R2
o5: t6:R3 $\leftarrow$ xor t1:R1 t5:R3
o7: t8:R0 $\leftarrow$ xor t0:R0 t6:R3
o9: out [t10:R0]
\end{lstlisting}
\end{tcolorbox}
\caption{\label{lst:unimasked4} Solution of the model in Figure~\ref{lst:unimasked}}
\end{figure}

\subsubsection{Mmpairs Constraints}
\label{ssec:mmconstraint}
The following constraint ensures that a pair of non-secret memory
operations in $Mmpairs$, are either not active or not subsequent
memory operations (\texttt{msubseq} constraint).
Constraint \texttt{msubseq} (defined in Section~\ref{sec:subseq}) is similar to
\texttt{subseq} but considers consecutive memory operations instead of
temporaries.

\begin{tcolorbox}[colback=\colbackcolor,colframe=\colframecolor, top=-5pt, bottom = -5pt,
  right = 0pt, left = 0pt]
\begin{lstlisting}[style = modelingstyle]
forall (o$_1$,o$_2$) in $Mmpairs$:
  a(o$_1$) $\land$ a(o$_2$) $\implies$
    ($\neg$msubseq(o$_1$,o$_2$) $\land$ $\neg$msubseq(o$_2$,o$_1$))
\end{lstlisting}
\end{tcolorbox}

\subsubsection{Mspairs Constraints}
\label{ssec:msconstraint}
Finally, the following constraint ensures that for each pair
$(\texttt{o}_s,\texttt{o}_{rs}) \in Mspairs$ a random memory
operation $\texttt{o}_r \in \texttt{o}_{rs}$ precedes the
secret-dependent memory operation $\texttt{o}_s$.

\begin{tcolorbox}[colback=\colbackcolor,colframe=\colframecolor,
    top=-5pt, bottom = -5pt,
    right = 0pt, left = 0pt]
\begin{lstlisting}[style = modelingstyle]
forall (o$_s$,o$_{rs}$) in $Mspairs$:
  exists o$_r$ in o$_{rs}$:
    a(o$_s$) $\implies$ (a(o$_r$) $\land$ msubseq(o$_r$,o$_s$))
  $\land$
  exists o$_r$ in o$_{rs}$:
    a(o$_s$) $\implies$ (a(o$_r$) $\land$ msubseq(o$_r$,o$_s$))
\end{lstlisting}
\end{tcolorbox}

This constraint works similarly as the equivalent register constraint,
where instead of register operations, we have memory operations.
In our example, we need to have memory spilling, i.e. store to the
stack, and then load from the stack (only one of the operations is
shown in Figure~\ref{lst:unimasked4}).

\subsubsection{Modeling \texttt{subseq}}
\label{sec:subseq}

To define the \texttt{subseq} constraint, we first define an auxiliary predicate
\texttt{is\_before} and a set of auxiliary problem variables \texttt{lk}.
Predicate \texttt{is\_before(t$_1$, t$_2$)} tells whether \texttt{t$_1$} is
assigned to the same register as \texttt{t$_2$} and \texttt{t$_1$}'s life range
ends (\texttt{le(t$_1$)}) before the beginning of the life range of
\texttt{t$_2$} (\texttt{ls(t$_2$)}).

\begin{tcolorbox}[colback=\colbackcolor,colframe=\colframecolor, top=-5pt, bottom = -5pt,
  right = 0pt, left = 0pt]
\begin{lstlisting}[style = modelingstyle]
  pred is_before(t$_1$,t$_2$): same_reg(t$_2$, t$_1$) $\land$
              (le(t$_2$) $\le$ ls(t$_1$))
\end{lstlisting}
\end{tcolorbox}

Variable \texttt{lk(t)} captures the end live cycle of the temporary
that occupied the same register as \texttt{t} (\texttt{r(t)}) right
before \texttt{t} was assigned.
If \texttt{t' = lk(t)}, then the values of \texttt{t} and
\texttt{t'} result in a transitional effect that may reveal
information to the attacker.

\begin{tcolorbox}[colback=\colbackcolor,colframe=\colframecolor,
    top=-5pt, bottom = -5pt,
    right = 0pt, left = 0pt]
\begin{lstlisting}[style = modelingstyle]
forall t in Temps: lk(t) = max(
  [ite(is_before(t$'$,t),le(t$'$),-1)
                  | forall t$'$ in Temps])
\end{lstlisting}
\end{tcolorbox}

Then, the definition of the \texttt{subseq} predicate is as follows:
\begin{tcolorbox}[colback=\colbackcolor,colframe=\colframecolor, top=-5pt, bottom = -5pt,
  right = 0pt, left = 0pt]
\begin{lstlisting}[style = modelingstyle]
pred subseq(t$_1$,t$_2$):
  samereg(t$_1$,t$_2$) $\land$ (lk(t$_2$) = le(t$_1$))
\end{lstlisting}
\end{tcolorbox}

\begin{theorem}[Subseq Constraint] The \texttt{subseq} constraint is true
  only for pairs of temporary variables that are subsequently assigned
  to the same register:

  \texttt{subseq}($t_1$,$t_2$) $\Longleftrightarrow$ $P = P'; t_1
  \leftarrow e_1; P''; t_2 \leftarrow e_2; P'''$ $\land$ $r(t_1) =
  r(t_2)$ $\land$ $\forall i \in P''~.~ i = t \leftarrow e \implies r(t) \neq r(t_1)$.
  \label{th:instrseq}
\end{theorem}
\begin{proof}
  $(\Leftarrow)$ Assume $P = P'; t_1 \leftarrow e_1; P''; t_2
  \leftarrow e_2; P'''$ $\land$ $r(t_1) = r(t_2)$ $\land$ $\forall i
  \in P''~.~ i=t = e \land r(t) \neq r(t_1)$.
  We consider all register assignments in $P$: $P = ...; t_i
  \leftarrow e_i; ...; t_1 \leftarrow e_2; ...; t_2 \leftarrow e_2;
  ...; t_j \leftarrow e_j...$; all these assignments are live because
  they appear in the final program.
  For all assignments $t_j$ following $t_i$ we have that $le(t_j) >
  ls(t_2)$, which implies that $\texttt{is\_before}(t_j,t_i) =
  \texttt{false}$, and thus all $t_j$ contribute with -1 to \texttt{max} in $lk(t_2)$.
  The same applies for all registers that are assigned to a
  different register, they contribute with -1 because
  $\texttt{is\_before}(t_j,t_i) =
  \texttt{false}$.
  Then, $lk(t_2) = max(le(t) | t \in \{t_{i_1},t_{i_2},..,t_1\})$,
  where all $\{t_{i_1},t_{i_2},..,t_1\}$ are assigned the same
  register, $r(t_2)$.
  Because these temporaries are assigned to the same register, their
  live ranges do not overlap (Equation~\ref{eq:nooverlap}),
  i.e. $\forall t,t' \in \{t_{i_1},t_{i_2},..,t_1\}~.~ ls(t) \ge
  le(t') ~\lor~ ls(t') \ge le(t)$.
  Because $t_1 \leftarrow e_1$ is scheduled last $\forall t \in
  \{t_{i_1},t_{i_2},..,t_{i_n},t_1\}~.~ ls(t_1) \ge le(t)$.
  Also, from Equation~\ref{eq:liveend}, $le(t_1) > ls(t_1)$.
  This implies that $\forall t \in \{t_{i_1},t_{i_2},..,t_{i_n}\}~.~
  le(t_1) > le(t)$, so we have $lk(t_2) = le(t_1)$ and $\forall t \in
  \{t_{i_1},t_{i_2},..,t_{i_n}\}~.~ lk(t_2) > le(t)$.
  Therefore only for $t_1$, \texttt{subseq}($t_1,t_2$) = \texttt{true}.
  
  $(\Rightarrow$) Assume \texttt{subseq}($t_1,t_2$).
  This implies that \texttt{samereg}($t_1,t_2$) $\land$ $lk(t_2) = le(t_1)$.
  Constraint \texttt{samereg}($t_1,t_2$) implies that $r(t_1) =
  r(t_2)$ and $l(t_1) \land l(t_2)$, which means that they appear in
  the final code, $P$, and are assigned to the same register.
  Because $lk(t_2) = le(t_1)$, $t_1$ is scheduled before $t_2$ or
  $P = P'; t_1 \leftarrow e_1; P''; t_2 \leftarrow e_2; P'''$.
  Now, we only need to prove that there is no other assignment of
  $r(t_1)$ in $P''$, i.e.\ $\forall i \in P''~.~ t \leftarrow e \land
  r(t) \neq r(t_1)$.
  If $\exists i \in P''~.~ t \leftarrow e \land r(t) = r(t_1)$,
  then, because live ranges do not overlap, $le(t) > le(t_1)$, which
  means that $lk(t_2) = le(t), \neq le(t_1)$, which is invalid.
  
\end{proof}

For the definition of \texttt{msubseq}, we define an auxiliary
predicate \texttt{is\_before\_mem} and auxiliary problem variables
\texttt{ok}.
Predicate \texttt{is\_before\_mem(o$_1$, o$_2$)} tells whether \texttt{o$_1$} is
scheduled before \texttt{o$_2$}.

\begin{tcolorbox}[colback=\colbackcolor,colframe=\colframecolor, top=-5pt, bottom = -5pt,
  right = 0pt, left = 0pt]
\begin{lstlisting}[style = modelingstyle]
pred is_before_mem(o$_1$,o$_2$):
    a(o$_1$) $\land$ (c(o$_1$) $\leq$ c(o$_2$))
\end{lstlisting}
\end{tcolorbox}
\noindent
In Figure~\ref{lst:unimasked4}, \texttt{is\_before\_mem(o$_4$, o$_3$)}
is \texttt{true}.

Variable \texttt{ok(o)} captures the issue cycle of memory operation
\texttt{o$'$ $\in$ MemOperations} that was issued before $o$.

\begin{tcolorbox}[colback=\colbackcolor,colframe=\colframecolor,
    top=-5pt, bottom = -5pt,
    right = 0pt, left = 0pt]
\begin{lstlisting}[style = modelingstyle]
forall o in MemOperations: ok(o) = max(
  [ite(is_before_mem(o$'$, o), c(o$'$), -1)
          | forall o$'$ in MemOperations])  
\end{lstlisting}
\end{tcolorbox}

Similar to predicate \texttt{subseq}, \texttt{msubseq} is as follows:

\begin{tcolorbox}[colback=\colbackcolor,colframe=\colframecolor,
    top=-5pt, bottom = -5pt,
    right = 0pt, left = 0pt]
\begin{lstlisting}[style = modelingstyle]
pred msubseq(o$_1$,o$_2$):
  a(o$_1$) $\land$ a(o$_2$) $\land$ ok(o$_2$) = c(o$_1$)
\end{lstlisting}
\end{tcolorbox}

\begin{theorem}[Msubsec Constraint] The \texttt{msubseq} constraint is true
  only for two instructions that are subsequently accessing the memory:
  \texttt{msubseq}($o_1$,$o_2$) $\Longleftrightarrow$ $P = P';o_1;
  P''; o_2; P'''$ $\land$ $\nexists o \in P''~.~ o = \texttt{mem}(e'',
  e_3)$, where $o_1$ and $o_2$ are memory operations, $o_1 =
  \texttt{mem}(e, e_1) $ and $o_2 = \texttt{mem}(e', e_2)$.
  \label{th:msubseq}
\end{theorem}
\begin{proof}
  Similar to Theorem~\ref{th:instrseq}.
\end{proof}

Theorem~\ref{th:proof} shows that \ac{\toolname} generates secure code for our
threat model.

%% To prove that \ac{\toolname} generates secure programs,

\begin{theorem}[Secure Modeling]
  A program P, generated by \ac{\toolname}, satisfies the
  \textit{leakage equivalence} condition in
  Definition~\ref{def:security_condition}.
  This means that given two input instances $\mathit{IN}$, $\mathit{IN'}$ that differ
  only with regards to the secret variables, $\mathit{IN}_{sec} \subseteq
  \mathit{IN}$, $\mathit{IN'}_{sec} \subseteq \mathit{IN'}$, the distributions of the leakages do
  not differ.
  \label{th:proof}
\end{theorem}
\begin{proof}
  We assume that the type-inference algorithm overapproximates the actual
  distribution of each variable.
  Then, we perform structural induction on the program $P$ to prove
  that security constraints we introduce lead to secure programs.
  The proof is available as supplementary material~\cite{appendix}.
\end{proof}

%% , we sketch a proof
%% that our model leads to programs that do not leak secret information
%% according to the leakage model in Equation~\ref{eq:lmodel}.

%% \subsubsection{Search}
%% \label{ssec:search}
%% In combinatorial problem solving, \textit{search} affects the
%% efficiency of the heuristic.
%% %
%% A constraint-based compiler backend is typically optimized to find the
%% optimal among all possible solutions.
%% %
%% However, our new constraints increase the complexity of the model and
%% reduce the number of available solutions.
%% %
%% A previously optimal solution may not satisfy the security
%% constraints.
%% %
%% At the same time, changing completely the search strategy does not
%% benefit the solving the total program.
%% %
%% For solving our problem, we introduce a small number of additional
%% branching decision for leading search to possible solution to the
%% secure program.

%% In particular the following branching strategy decides whether a
%% secret temporary that is part of $Spairs$ will be active or inactive.
%% %
%% This branching strategy enables early decision on whether the secret
%% temporary will actually be active or not.

%% \begin{tcolorbox}[colback=\colbackcolor,colframe=\colframecolor,
%%     top=-5pt, bottom = -5pt,
%%     right = 0pt, left = 0pt]
%% \begin{lstlisting}[style = modelingstyle]
%% for ($t_s$,$t_{rs}$) in $Spairs$:
%%   ls.append(l(t$_s$))
%% branch(ls);
%% \end{lstlisting}
%% \end{tcolorbox}

\subsection{Solving Enhancements}
\label{sec:search}

Large problems in combinatorial solving can quickly become
difficult to handle due to state-space explosion.
A solution to this problem is structural decomposition of the
problem into subproblems.
In code generation, a natural structural decomposition scheme
consists of splitting the problem into basic
blocks~\cite{lozano_combinatorial_2019}.
However, \ac{\toolname}'s security
analysis~\cite{wang_mitigating_2019-1} requires linearized code that
corresponds to  one large basic block.
There are already approaches on splitting large code blocks into
smaller artificial code blocks for improving the scalability of the
solver~\cite{lozano_combinatorial_2019}.
Typically, in decomposition schemes, the solver first solves each
partial solution (basic blocks) and then composes a full solution
consisting of the partial solutions.
However, this solution becomes challenging with the addition of
security constraints that relate different parts of the code,
introducing new inter-block dependencies.
%
%% To overcome this problem, we first perform structural decomposition
%% using one of two methods, \textit{spectral clustering} or linear
%% code splitting~\cite{lozano_combinatorial_2019}.
%
%% Typically, the solver first solves each partial solution (basic
%% block) and then composes a full solution consisting of the partial
%% solutions.
%
These dependencies may lead to conflicts between the partial
solutions resulting in the rejection of the full solution.
%
%% \roberto{I found this part a bit confusing}
%
To deal with this problem, \ac{\toolname} propagates only part of
the partial solutions, leaving some parts of the full
solution unsolved.
%
%% These unsolved parts correspond to the connection points between
%% the code blocks.
%
In particular, \ac{\toolname} does not propagate the register
assignments to temporaries that correspond to earliest and latest
assigned hardware registers in each basic block, as well as
their corresponding issue cycles.
%
%% The reason for this is that the latest temporary assigned to a
%% hardware register may conflict with the first temporary assigned to
%% the same hardware register in the following piece of code.
%
Subsequently, \ac{\toolname} solves the unsolved parts as part of
the full problem.

The second main enhancement to the solving procedure concerns
the final step of the solving process.
In \ac{\toolname} we make use of \ac{LNS}~\cite{shaw_using_1998},
a form of local search for constraint programming.
In particular, at the end of the decomposition phase,
\ac{\toolname} uses the best found solution to perform local search
and locate better solutions.

\section{Evaluation}
\label{sec:eval}
This section evaluates \ac{\toolname} focusing on three axes:
%% This section describes the evaluation of \ac{\toolname}.
%% %
%% For the evaluation of \ac{\toolname}, we pose the following research
%% questions:
\begin{description}
\item[Performance Overhead]
  What is the overhead in execution time for the generated code using
  \ac{\toolname}?
  Here, we want to evaluate the introduced overhead of secure
  solutions compared to optimized but insecure solutions.
  To do that, we compare the best known
  solution~\cite{lozano_combinatorial_2019} with our approach
  \ac{\toolname}.
\item[Performance Improvement] What is the improvement in execution
  time of the generated code over non-optimized code and other
  \ac{TBL}-secure approaches?
  Here, we compare our results with LLVM-3.8 with no optimization
  (-O0) and the work by
  \citeauthor{wang_mitigating_2019-1}~\cite{wang_mitigating_2019-1}.
\item[Compilation Overhead]
  What is the overhead in compilation time using \ac{\toolname}?
  Here, we want to evaluate the introduced compilation overhead of
  secure solutions compared to insecure solutions.
  To do that, we compare the compilation time for
  retrieving the best known solution~\cite{lozano_combinatorial_2019}
  with \ac{\toolname}'s compilation time.
\end{description}

\subsection{Preliminaries}
The following sections describe the implementation details and the
experimental setup of the evaluation of \ac{\toolname}.
The implementation of \ac{\toolname} and the experiments and
benchmarks for the evaluation are available at
\url{https://github.com/romits800/seccon_experiments.git}.

\subsubsection{Implementation Details}
\ac{\toolname} is implemented as an extension of
Unison\footnote{Unison:
    \url{http://unison-code.github.io/}}~\cite{lozano_combinatorial_2019},
a constraint-based compiler backend that uses \ac{CP} to optimize
software functions with regards to code size and execution time.
In particular, Unison combines two low-level optimizations,
instructions scheduling and register allocation, and achieves
optimizing medium-size functions with improvement compared to LLVM.
Unison uses two global constraints for modeling the backend
transformations; 1) the \textit{geometric packing constraint} for
register allocation and 2) the \textit{cumulative} constraint for
instruction scheduling.
The type-inference implementation is written in Haskell and is based
on \citeauthor{wang_mitigating_2019-1}~\cite{wang_mitigating_2019-1}
with precision improvements (see supplementary material~\cite{appendix}).

\subsubsection{Experimental Setup}

All experiments run on
an Intel%
\textsuperscript{\textregistered}%
Core\texttrademark i9-9920X processor at 3.50GHz with 64GB of RAM
running Debian GNU/Linux 10 (buster).
We use LLVM-3.8 as the front-end compiler for these experiments.
To preserve the high-level security properties of the compiled
programs, we apply only one optimization, register promotion,
(\texttt{-mem2reg} in LLVM), which lifts program variables from the
stack to registers.
We evaluate our method on two architectures: \thumb, targeting
processor ARM Cortex M0, a highly predictable processor targeting
small embedded devices; and \mips, a widely-used embedded
architecture.

We implemented the constraint model both as part of the specialized
Gecode~\cite{Gecode2020} constraint model and the
Minizinc~\cite{nethercote_minizinc_2007} model that Unison provides.
The Minizinc model allows for solving the problem using multiple
solvers.
In total, we tried four solvers, Chuffed v0.10.3~\cite{Chu2011},
OR-Tools~\cite{ortools}, Elsie Geas\footnote{Elsie Geas:
  \url{https://bitbucket.org/gkgange/geas/src/master/}}, and the
specialized model written in Gecode v6.2.
We ran the former three solvers activating the \textit{free-search}
option.
For the specialized model in Gecode, apart from the security model,
\ac{\toolname} includes the modified search enhancements that we
describe in Section~\ref{sec:search}.
Among all these solvers, Gecode and Chuffed performed best.
None of them was able to solve all the problems but together they
could solve most of the problems.
In the smaller benchmarks, P0-P6, we run a portfolio solver
including Gecode and Chuffed.
For the larger benchmarks, we run every solver separately for reducing
the risk of out-of-memory errors when running both solvers in
parallel.
The presented results are the result of five runs for \ac{\toolname}
and Unison, whereas for the calculation of the execution time for LLVM
-O0, we run the compilation 1000 times to account for possible
fluctuations in the compilation time on the test machine.

\subsubsection{Benchmarks}

To evaluate our approach, we use a set of small benchmark programs, up
to 100 lines of C code and one program exceeding 900 lines of C code.
Table~\ref{tab:bench_description} provides a description of these
benchmarks, including the number of lines of code (LoC), and the
program variables, i.e.\ the input variables ($\mathit{IN}$) and the
number of secret ($\mathit{IN}_{sec}$), public ($\mathit{IN}_{pub}$),
and random ($\mathit{IN}_{rand}$) input variables.
Benchmarks P1 to P6 and P8 to P11 were made available by \citeauthor{wang_mitigating_2019-1}%
\footnote{FSE19 tool:
    \url{https://github.com/bobowang2333/FSE19}}~\cite{wang_mitigating_2019-1},
whereas P0 and P7 are implemented by the authors of this paper.
These benchmark programs constitute different masked implementations
from previous work and are linearized.
\citeauthor{wang_mitigating_2019-1}~\cite{wang_mitigating_2019-1}
  use a larger number of benchmarks to evaluate their approach.
However, our approach depends on an combinatorial optimizing compiler,
Unison, which scales to up to medium size functions, namely, up to
approximately 200 intermediate instructions for ARM Cortex M0 and
\mips architectures~\cite{lozano_combinatorial_2019}.
In addition to this, \ac{\toolname} adds additional constraints that
increase the complexity of the model (see Section~\ref{ssec:rq3}).
Therefore, we selected the smallest benchmarks for our experiments.
As a future work, we plan to investigate non-linearized
implementations, but this comes at the expense of analysis precision
and potentially increased performance overhead.
%% at the cost on precision.

\begin{table}[h]
  \centering
  \begin{tabular}{|l|l|r|r|r|r|}
    \hline
    \multirow{2}{*}{Prg} & \multirow{2}{*}{Description}     & \multirow{2}{*}{LoC}  & \multicolumn{3}{c|}{Input Variables (IN)} \\\cline{4-6}
    &   &  & pub & sec & rand \\\hline
    P0  & Xor (Listing~\ref{lst:masked})                         & 5    & 1  & 1 &  1  \\
    P1  & AES Shift Rows~\cite{bayrak_sleuth_2013}               & 11   & 0  & 2 &  2  \\
    P2  & Messerges Boolean~\cite{bayrak_sleuth_2013}            & 12   & 0  & 1 &  2  \\
    P3  & Goubin Boolean~\cite{bayrak_sleuth_2013}               & 12   & 0  & 1 &  2  \\
    P4  & SecMultOpt\_wires\_1~\cite{rivain_provably_2010}       & 25   & 1  & 1 &  3  \\
    P5  & SecMult\_wires\_1~\cite{rivain_provably_2010}          & 25   & 1  & 1 &  3  \\
    P6  & SecMultLinear\_wires\_1~\cite{rivain_provably_2010}    & 32   & 1  & 1 &  3  \\
    P7  & Whitening~\cite{bayrak_sleuth_2013}                    & 58   & 16 & 16&  16 \\
    P8  & CPRR13-lut\_wires\_1~\cite{coron_higher-order_2014}    & 81   & 1  & 1 &  7  \\
    P9  & CPRR13-OptLUT\_wires\_1~\cite{coron_higher-order_2014} & 84   & 1  & 1 &  7  \\
    P10 & CPRR13-1\_wires\_1~\cite{coron_higher-order_2014}      &104   & 1  & 1 &  7  \\
    P11 & KS\_transitions\_1~\cite{barthe_verified_2015-1}       &964   & 1  & 16 &  32  \\\hline
    %% P11  & KS\_wires~\cite{barthe_verified_2015-1}               &1130  & 1  & 16 &  32  \\\hline
  \end{tabular}
  \caption{\label{tab:bench_description} Benchmark Description}
\end{table}

\subsection{Optimality Overhead}
\label{ssec:rq1}

\ac{\toolname} builds on a constraint-based compiler backend to
generate a program that satisfies security constraints for software
masking.
This means that our approach might compromise some of the code quality
of the non-mitigated optimized code to mitigate the software masking
leaks.
To evaluate the overhead of our method compared to non-secure
optimization, we compare the execution time of the optimized solution
(optimal or suboptimal solution) that
Unison~\cite{lozano_combinatorial_2019} generates compared with
\ac{\toolname}'s optimized and \ac{TBL}-secure code.
The overhead is computed as $(\mathit{cycles}(\mathit{\toolname}) -
\mathit{cycles}(\mathit{Unison})) / \mathit{cycles}(\mathit{Unison})$.

Table~\ref{tab:secunison_vs_unison} shows the mean execution time for
each of the benchmark programs and architectures.
In particular, for each of the architectures, we compare the execution
time in number of cycles of the solution that Unison produces against
\ac{\toolname}'s solution.
The final column shows the overhead of \ac{\toolname} compared to
Unison.

The results show zero overhead for \mips, and a maximum \largestoh
overhead in ARM Cortex M0.
The zero overhead for most of the benchmarks shows that the Pareto
front of optimal solutions synthesized by Unison includes code
variants that are secure.
This result is in agreement with previous
work~\cite{tsoupidi2021constraint}, which shows the existence of
multiple optimal (or best found) solutions.
For ARM Cortex M0, programs P6 and P10 have a non-zero
positive overhead.
The observed overhead in ARM Cortex M0 is due to three main reasons:
1) the mitigation itself that may require the introduction of
redundant operations in the generated code, 2) the scalability issue
that appears in larger functions due to the addition of new security
constraints in the order of $|Temps|^2$, and 3) the decomposition
mode that may fail to compose solutions (Section~\ref{sec:search}).
Programs P8 and P9 show a slight improvement.
This improvement is due to the introduction of \ac{LNS} at the end
of the solving stage (see Section~\ref{sec:search}).
%
%% This leads to an increase in the search space and hinders the
%% constraint model to locate better solutions.
The last benchmark program, P11, demonstrates the scalability limits
of our approach.
The operating system terminates the solving process because the
process attempts to allocate more than the available memory
(out-of-memory error).

To summarize, \ac{\toolname} does not introduce significant overhead
over the non-secure optimized solution that Unison generates.
This means that in most cases, there is space for generating secure
code without affecting the quality of the generated code.

\begin{table}[h]
  \centering
  \begin{tabular}{|l|r|r|g|r|r|g|}
    \hline
    \multirow{2}{*}{Prg}          & \multicolumn{3}{c|}{ARM Cortex M0}       & \multicolumn{3}{c|}{\mips}\\\hhline{~------}
    & \cite{lozano_combinatorial_2019}  & SCG& \soverhead (\%) & \cite{lozano_combinatorial_2019} & SCG & \soverhead (\%) \\\hline
P0	&	4	&	4	&	0	&	3	&	3	&	0\\
P1	&	5	&	5	&	0	&	4	&	4	&	0\\
P2	&	8	&	8	&	0	&	7	&	7	&	0\\
P3	&	11	&	11	&	0	&	9	&	9	&	0\\
P4	&	25	&	25	&	0	&	76	&	76	&	0\\
P5	&	25	&	25	&	0	&	76	&	76	&	0\\
P6	&	24	&	25	&	4	&	74	&	74	&	0\\
P7	&	120	&	120	&	0	&	184	&	184	&	0\\
P8	&	81	&	80	&	-1	&	152	&	152	&	0\\
P9	&	86	&	85	&	-1	&	152	&	152	&	0\\
P10	&	90	&	96	&	7	&	282	&	282	&	0\\
P11	&	1558	&	OM	&	-	&	1335	&	OM	&	-\\
\hline
  \end{tabular}                                    
  \caption{\label{tab:secunison_vs_unison} Optimal solution by Unison
    and \ac{\toolname} (SCG) in cycles; \soverhead stands for overhead; OM
    stands for \textit{out of memory}}
\end{table}

\subsection{Execution-time Improvement}
\label{ssec:rq2}

To evaluate the execution-time speedup of our approach, we
compare \ac{\toolname} with the code generated by LLVM without
optimizations (-O0).
We also compare \ac{\toolname} with the work by
\citeauthor{wang_mitigating_2019-1}~\cite{wang_mitigating_2019-1}.
\citeauthor{wang_mitigating_2019-1} identify and mitigate \ac{ROT}
leaks on non-optimized code from LLVM 3.6.
This is a common approach by different security mitigations, because
compilation passes may violate the security properties of a program.
During their mitigation, \citeauthor{wang_mitigating_2019-1} may
remove unused code~\cite{wang_mitigating_2019-1}, which reduces the
overhead.
%
%% However, non-optimized code leads to
%% high performance overhead.
%% \footnote{ From preliminary experiments that we performed
%%   \ac{\toolname} is better than LLVM with -O0, with an improvement
%%   ranging from 75\% to 6x speedup}.
%% \roberto{maybe this could be moved to the intro (e.g.~after \emph{However, unoptimized code is highly inefficient.}) to make that clear from the beginning.}

We compare \ac{\toolname} with the approach by
\citeauthor{wang_mitigating_2019-1}~\cite{wang_mitigating_2019-1} for
three main reasons, 1) their tool is available freely, 2) they propose
an architecture-agnostic approach that applies to both \mips and \thumb,
and 3) they mitigate transitional effect caused by register reuse, a
subset of our mitigation.
Table~\ref{tab:secunison} compares the execution time in number of
cycles (based on a LLVM-derived cost model) of LLVM, the mitigated code by
\citeauthor{wang_mitigating_2019-1}~\cite{wang_mitigating_2019-1} and
\ac{\toolname}, for each of the programs and architectures.
Speedup is computed as $\mathit{cycles}(\mathit{\toolname}) /
\mathit{cycles}(\mathit{LLVMO0})$.

For ARM Cortex M0, the speedup ranges from \lowestsuthumb for
\lowestsuthumbp to \largestsuthumb for \largestsuthumbp and a
geometric mean of 3.0 speedup.
We notice that for the smaller benchmarks, \ac{\toolname} achieves
increased improvement over the baseline, whereas for the largest
benchmarks P7-P10, the improvement is smaller, but still significant.
%
%% These results are consistent with the results in
%% Section~\ref{ssec:rq1}, where the larger benchmarks, P7-P9, are not
%% able to reach the same quality level as in the absence of the security
%% constraints.
%% %
The main reason for this, is that the increased size of the program
under analysis reduces the ability of the solver to find optimal
solutions.

For \mips, the improvement ranges from \lowestsumips to
\largestsumips speedup and a geometric mean of 3.2 speedup.
The improvement is larger for smaller benchmarks due to the large
overhead of \texttt{load} and \texttt{store} instructions that are
present in the absence of optimizations in the baseline.
In contrast to the non-optimized code, the code generated by
\ac{\toolname} reduces memory spilling.
In particular, the generic cost model for \mips that we use (derived
from LLVM) has an one cycle overhead compared to linear instructions.
For larger programs, P4-P10, the speedup is smaller but still
significant.

This experiment shows that for both architectures \ac{\toolname}
achieves improvement ranging from 75\% up to a speedup of 8 with
geometric-mean speedups 3.0 and 3.2 for ARM Cortex M0 and \mips,
respectively.
Although not completely comparable with \ac{\toolname} because of the
use of different benchmarks and mitigations,
\citeauthor{vu_reconciling_2021-1} show an improvement over
non-optimized code (-O0) that ranges from 20\% to a speedup of 12.6,
with a geometric mean of 2.8~\cite{vu_reconciling_2021-1}.
Compared to the approach by \citeauthor{wang_mitigating_2019-1}, the
speedup that \ac{\toolname} achieves ranges from 1.24 (24\%) to 6.5
for ARM Cortex M0 and from 1.36 (36\%) to 7.6 for \mips.
The geometric-mean speedups are 2.6 for ARM Cortex M0 and 2.9 for
\mips.
%% \todo{After adding the part about -O0, \cite{vu_reconciling_2021-1}
%%   reports 1.2 to 12.6 speedup and 2.8 geometric mean}

To summarize, for both \mips and ARM Cortex M0, \ac{\toolname}
improves the non-optimized LLVM code.
We notice large improvements for both \mips and ARM Cortex M0 ranging
from \lowestsumips to \largestsumips speedup.
\ac{\toolname} generates also improved code compared to the work by
\citeauthor{wang_mitigating_2019-1}~\cite{wang_mitigating_2019-1}.

\begin{table}[h]
  \centering
  \begin{tabular}{|l|r|r|r|g|r|r|r|g|}
    \hline
    \multirow{2}{*}{Prg}             & \multicolumn{4}{c|}{ARM Cortex M0}       & \multicolumn{4}{c|}{\mips}\\\hhline{~--------}
    & O0 & \cite{wang_mitigating_2019-1}  & SCG & \sspeedup & O0 & \cite{wang_mitigating_2019-1} & SCG & \sspeedup \\\hline
P0	&	13	&	13	&	4	&	3.25	&	19	&	23	&	3	&	6.33\\
P1	&	29	&	22	&	5	&	5.80	&	33	&	21	&	4	&	8.25\\
P2	&	55	&	52	&	8	&	6.88	&	43	&	43	&	7	&	6.14\\
P3	&	32	&	33	&	11	&	2.91	&	47	&	47	&	9	&	5.22\\
P4	&	61	&	61	&	25	&	2.44	&	139	&	139	&	76	&	1.83\\
P5	&	58	&	58	&	25	&	2.32	&	133	&	133	&	76	&	1.75\\
P6	&	78	&	45	&	25	&	3.12	&	189	&	188	&	74	&	2.55\\
P7	&	313	&	465	&	120	&	2.61	&	382	&	430	&	184	&	2.08\\
P8	&	182	&	106	&	80	&	2.27	&	371	&	253	&	152	&	2.44\\
P9	&	187	&	181	&	85	&	2.20	&	371	&	371	&	152	&	2.44\\
P10	&	218	&	119	&	96	&	2.27	&	593	&	383	&	282	&	2.10\\
P11	&	4100	&	3864	&	OM	&	-	&	3688	&	3237	&	OM	&	-\\
\hline
  \end{tabular}                                    
  \caption{\label{tab:secunison} Execution-time comparison between the
    non-optimized baseline and \ac{\toolname} (SCG); \sspeedup is the
    speedup of \ac{\toolname} with LLVM with -O0 as baseline; OM
    stands for \textit{out of memory}}
\end{table}

\subsection{Compilation Overhead}
\label{ssec:rq3}

%% \roberto{The O0 columns in Table~\ref{tab:secunison_vs_unison_compilation} give
%%   very little information. I would simply remove them from the table and
%%   summarize them in the running text instead. Then define Sd =
%%   $\mathit{cycles}(\mathit{\toolname}) / \mathit{cycles}(\mathit{Unison})$ }

To evaluate the compilation overhead of our approach, we compare
\ac{\toolname} with Unison~\cite{lozano_combinatorial_2019} and
non-optimized LLVM.
The main reason for the compilation overhead of \ac{\toolname}
compared to LLVM is the combinatorial nature of the backend compiler.
Compared to Unison, \ac{\toolname} introduces compilation overhead due
to the security constraints among temporaries and operations in the
combinatorial model.
In particular, the \texttt{subseq} constraint introduces a large
number of constraints and variables that are in the order of
$|Temp|^2$.
The constraints between memory operations (\texttt{msubseq}) are
typically fewer because memory operations are a subset of all
operations.
In general, the actual overhead depends on the program logic and the
security policy.
The compilation slowdown is computed as
$\mathit{comp\_time}(\mathit{\toolname}) /
\mathit{comp\_time}(\mathit{Unison})$.
%% \roberto{confusing: this is not what ``Sd'' gives in the table, right?}.
% \romy{I updated the cycles to comp\_time I guess this is what you meant?

Table~\ref{tab:secunison_vs_unison_compilation} compares the
compilation time of \ac{\toolname} and Unison.
The last column for each architecture in
Table~\ref{tab:secunison_vs_unison_compilation} presents the slowdown
of \ac{\toolname} compared to Unison.
In \mips, we can see an increasing overhead in the compilation time of
\ac{\toolname} compared to Unison with the increase of the function
size.
The largest compilation overhead is for P10 and corresponds to
\mipsmaxcohunison slowdown compared to Unison.
The compilation time for non-optimized LLVM ranges from 0.01 to 0.04
seconds.
Comparing \ac{\toolname} with LLVM, the slowdown ranges from 98 for P0
to \mipsmaxcoh for P10 (the detailed results are excluded from
Table~\ref{tab:secunison_vs_unison_compilation} due to lack of space).

In the case of ARM Cortex M0, we observe a similar trend.
We observe the largest slowdown for P9 which corresponds to
\thumbmaxcohunison slowdown.
However, the compilation time increases faster than for
\mips.
Compared with LLVM, \ac{\toolname} results in a slowdown that ranges
from 27 for P0 to \thumbmaxcoh for P10 (does not appear in
Table~\ref{tab:secunison_vs_unison_compilation}).
%% Program P4 is a special case because compilation with \ac{\toolname} is
%% faster than with Unison.
%% %
%% The reason for this is that while Unison cannot prove the optimality
%% of the code in the portfolio mode, it continues searching.
%% %
%% However, \ac{\toolname} is able to prove that the final solution is
%% optimal and, thus, terminates the compilation process.
%% %
%% For P5, we have the opposite case, where \ac{\toolname} continues
%% searching for better solution.
%% %
%% For the rest of the benchmarks, we have similar pattern as for \mips,
%% with the different that the overhead for ARM Cortex M0 is high, up to
%% \thumbmaxcoh, but lower than for \mips.
%% % 
The main reasons for the observable difference between the two
architectures are 1) the \thumb architecture is more
constrained\footnote{ARM Cortex M0 has fewer general-purpose registers
  than MIPS32 and includes two-address instructions, which restrict
  register allocation.}
%% 2) there are internal
%% timeouts in Unison and by inheritance to \ac{\toolname},
and 2)
interestingly, most instances for \mips are solved quickly by Chuffed,
whereas most instances for ARM Cortex M0 are only solved by Gecode.

To summarize, the compilation time for \ac{\toolname} is multiple
times slower than Unison because of the introduction of security
constraints.
\ac{\toolname} is orders of magnitude slower than LLVM.
Therefore, we believe that \ac{\toolname} is mostly suitable for
compiling small cryptographic kernels that are both critical for the
performance and the \ac{PSC} security, such as secure field
multiplication for AES~\cite{rivain_provably_2010}.
%% We observe a difference between \mips and ARM Cortex M0, where
%% the former increases in a regular way, whereas 

%

%% \begin{table}[h]
%%   \centering
%%       {%% TODO(Romy): todo or not todo?
%%         %% \setlength{\tabcolsep}{5.5pt}
%%   \begin{tabular}{|l|r|r|r|g|r|r|r|g|}
%%     \hline
%%     \multirow{2}{*}{Prg}          & \multicolumn{4}{c|}{ARM Cortex M0}       & \multicolumn{4}{c|}{\mips}\\\hhline{~--------}
%%     & O0 & \cite{lozano_combinatorial_2019}  & \toolnameshort & \sslowdown & O0 &\cite{lozano_combinatorial_2019} & \toolnameshort & \sslowdown \\\hline
%% P0	&	0.01	&	0.14	&	0.27	&	27.0	&	0.01	&	0.39	&	0.98	&	98.0\\
%% P1	&	0.01	&	0.14	&	0.32	&	32.0	&	0.01	&	0.44	&	1.3	&	0.1K\\
%% P2	&	0.01	&	0.28	&	1.0	&	0.1K	&	0.01	&	0.59	&	2.8	&	0.3K\\
%% P3	&	0.01	&	9.8	&	34.7	&	3.5K	&	0.01	&	0.69	&	3.8	&	0.4K\\
%% P4	&	0.01	&	0.7K	&	1K	&	0.1M	&	0.01	&	1.0	&	8.5	&	0.9K\\
%% P5	&	0.01	&	0.9K	&	1K	&	0.1M	&	0.01	&	1.0	&	8.5	&	0.8K\\
%% P6	&	0.01	&	63.5	&	0.4K	&	44.5K	&	0.01	&	1.1	&	10.1	&	1.0K\\
%% P7	&	0.01	&	0.9K	&	2K	&	0.2M	&	0.01	&	47.2	&	1K	&	0.1M\\
%% P8	&	0.01	&	0.1K	&	4K	&	0.4M	&	0.01	&	42.6	&	2K	&	0.2M\\
%% P9	&	0.01	&	0.3K	&	6K	&	0.6M	&	0.01	&	22.0	&	1K	&	0.1M\\
%% P10	&	0.04	&	7K	&	OM	&	-	&	0.04	&	\textit{52K}	&	OM	&	-\\
%% \hline
%%   \end{tabular}
%%   }
%%   \caption{\label{tab:secunison_vs_unison_compilation} Compilation
%%     overhead for \ac{\toolname} (SCG) compared to Baseline (LLVM -O0) in
%%     seconds; \sslowdown stands for slowdown; OM stands for ``out of
%%     memory''; numbers in \textit{italic} denote the use of
%%     swap memory}
%% \end{table}
%% }

\begin{table}[h]
  \centering
      {%% \setlength{\tabcolsep}{5.5pt}
  \begin{tabular}{|l|r|r|g|r|r|g|}
    \hline
    \multirow{2}{*}{Prg}          & \multicolumn{3}{c|}{ARM Cortex M0}       & \multicolumn{3}{c|}{\mips}\\\hhline{~------}
    & \cite{lozano_combinatorial_2019}  & \toolnameshort & \sslowdown &\cite{lozano_combinatorial_2019} & \toolnameshort & \sslowdown \\\hline
P0	&	0.14	&	0.27	&	1.9	&	0.39	&	0.98	&	2.5\\
P1	&	0.14	&	0.32	&	2.3	&	0.44	&	1.3	&	3.0\\
P2	&	0.28	&	1.0	&	3.6	&	0.59	&	2.8	&	4.8\\
P3	&	9.8	&	34.7	&	3.5	&	0.69	&	3.8	&	5.4\\
P4	&	0.7K	&	1K	&	1.4	&	1.0	&	8.5	&	8.3\\
P5	&	0.9K	&	1K	&	1.2	&	1.0	&	8.5	&	8.3\\
P6	&	63.5	&	0.4K	&	7.0	&	1.1	&	10.1	&	9.2\\
P7	&	3K	&	11K	&	3.3	&	6.3	&	0.1K	&	17.4\\
P8	&	0.9K	&	2K	&	2.5	&	47.2	&	1K	&	23.7\\
P9	&	0.1K	&	4K	&	37.9	&	42.6	&	2K	&	37.4\\
P10	&	0.3K	&	6K	&	19.1	&	22.0	&	1K	&	57.3\\
P11	&	7K	&	OM	&	-	&	\textit{52K}	&	OM	&	-\\
\hline
  \end{tabular}
  }
  \caption{\label{tab:secunison_vs_unison_compilation} Compilation
    overhead for \ac{\toolname} (SCG) compared to Baseline (Unison) in
    seconds; \sslowdown stands for slowdown; OM stands for \textit{out
      of memory}; numbers in \textit{italic} denote the use of swap
    memory}
\end{table}

\subsection{Threat to Validity}
Our model considers the \ac{HD} leakage model and generates code that
mitigates these leakages.
The security guaranties for our model depend on the \ac{HD} leakage
model.
The \ac{HD} model has been used both for designing
defenses~\cite{wang_mitigating_2019-1} and
attacks~\cite{brier_correlation_2004}.
However, the \ac{HD} model does not express precisely the actual
leakage model for some devices~\cite{oren_algebraic_2012}.
Moreover, an \ac{HD}-based mitigation at the assembly level may not
hold in the presence of advance microarchitectural features, such as
out-of-order execution and write buffers.
%% }
%% However, it is not certain that the removed vulnerabilities are
%% observable by an attacker on a real device and that the mitigated
%% program does not lead to further leakages in the actual hardware.
%
In addition to this, \ac{\toolname} does not handle transitional
effects through value interaction in the pipeline stage registers and
in the memory.
We leave further improvement of the hardware model as a future work.

\ac{\toolname} is not a verified compiler approach like
CompCert~\cite{leroy_formally_2009}.
Unison, the constraint-based backend that \ac{\toolname} depends on is
based on a formal model that implements standard optimizations but the
external solvers and the tool implementation are not verified.
Verification of constraint solvers is an active topic of
research~\cite{gocht2022auditable}.

\section{Related Work}

The following sections discuss the related work, with regards to
mitigations against side-channel attacks, combinatorial compilation
approaches, and mitigations against \acp{TBL}.
\citeauthor{athanasiou_automatic_2020-1} consider two types of
\ac{PSC} leakage sources, \acf{VBL} and \acf{TBL}.
\ac{VBL} occur due to the absence or compiler-induced removal of
masking.
As we have seen, \ac{TBL}, is a result of low-level
microarchitectural features such as register reuse, memory
overwrite, or interactions between values in the hardware.
In the following, we will use these two terms to describe
different mitigations.

\begin{table}
  \centering
  % todo or not todo
  \setlength{\tabcolsep}{5.5pt}
  \begin{tabular}{|l|m{1.2cm}|l|l|l|l|c|}
    \hline
    Pub. & Mitigation & Transf. & InL & OutL & \acs{ML}& Avail.\\\hline
    \rowcolor{gray!30}
    \cite{eldib_synthesis_2014-2}       &\acs{VBL}                    & \acs{FE}, \acs{ME}& \acs{DSL} & - & Custom & \cross \\
    \cite{bond_vale_2017-1}             &\acs{TSC}, \acs{MS}, \acs{RS} & - & \acs{DSL} & ASM & Custom &\fullcheckmark\\
    \rowcolor{gray!30}
    \cite{almeida_jasmin_2017}          &\acs{TSC}, \acs{MS}           & - & \acs{DSL} & ASM & Custom &\fullcheckmark \\
    \cite{zinzindohoue_hacl_2017-1}     &\acs{TSC}, \acs{MS}           & - & \acs{DSL} & C & F$_{low}$&\fullcheckmark\\
    \rowcolor{gray!30}
    \cite{cauligi_fact_2017}            &\acs{TSC}                    &\acs{ME} & \acs{DSL} & C & Custom & \fullcheckmark \\ %% https://github.com/PLSysSec/FaCT
    \cite{papagiannopoulos_mind_2017}   &\acs{TBL}                    &\acs{BE}  & AVR   & AVR & Binary & \fullcheckmark\\
    \rowcolor{gray!30}
    \cite{besson_information-flow_2019} &\acs{IFL}                    &\acs{BE} & C      & ASM & CompCert&{\color{orange!60!black}\faQuestion} \\
    \cite{wang_mitigating_2019-1}       &\acs{TBL}                    &\acs{BE} & C, C++ & ASM & LLVM &\fullcheckmark\\
    \rowcolor{gray!30}
    \cite{athanasiou_automatic_2020-1}  &\acs{TBL}                    &-        & ARM & ARM & Binary &\halfcheckmark\\
    \cite{vu_reconciling_2021-1}        &\acs{VBL}, \acs{TSC}, \acs{FI} & ALL & C, C++ & ASM & LLVM & \cross\\
    \rowcolor{gray!30}
    \cite{shelton_rosita_2021-1}        &\acs{TBL}                    &-        & ARM & ARM & Binary & \fullcheckmark\\
    \hline
    SCG                           &\acs{TBL}                    &\acs{ME}, \acs{BE} & C, C++ & ASM & LLVM &\fullcheckmark\\
    \hline
  \end{tabular}
  \caption{\label{tbl:rel} Mitigation approaches against side-channel attacks; SCG stands of \ac{\toolname}, \acs{FE}, \acs{ME}, \acs{BE} stands for front end, middle end, and back end, respectively; ASM stands for assembly}
\end{table}

\subsubsection*{Optimized Secure Compilation}

General purpose optimizing compilers perform transformations that may
invalidate high-level security mitigations or introduce security
flaws~\cite{dsilva_correctness-security_2015-1}.
However, performance is important for most security applications,
especially those operating on resource-restricted devices.
Table~\ref{tbl:rel} presents a non-exhaustive list of related work
that present compiler-based or binary-rewriting approaches against
side channel attacks.
For each publication (Publication), Table~\ref{tbl:rel}, shows
the mitigations of each approach (Mitigation), the
compiler level that each approach perform the mitigation (Transformation),
the input language (InL), the output language (OutL),
the \ac{ML} of each approach that is either a compiler or binary.
The last column (Avail.) denotes with \cross that the artifact is not
available, with \fullcheckmark that the artifact is available, with
\halfcheckmark that part of the artifact is available, and finally,
with {\color{orange!60!black}\faQuestion} where it is not clear
whether the artifact is available or not.

Multiple approaches present compiler-based mitigations against
\acp{TSC}~\cite{bond_vale_2017-1,almeida_jasmin_2017,
  zinzindohoue_hacl_2017-1,cauligi_fact_2017,vu_reconciling_2021-1},
proof of \ac{MS}~\cite{bond_vale_2017-1,almeida_jasmin_2017,
  zinzindohoue_hacl_2017-1}, or \ac{RS}~\cite{bond_vale_2017-1}.
In contrast, \ac{\toolname} targets \ac{PSC} attacks. 
%% Jasmin~\cite{almeida_jasmin_2017} is a framework for security-aware
%% and high-speed cryptographic implementation.
%% %
%% The framework includes a verified compiler that translates programs
%% from the Jasmin programming language to assembly language.
%% %
%% However, Jasmin focuses currently on constant-time programming and
%% does not provide support against power-side channels.
%% %
%% In addition to this, Jasmin requires writing programs to the Jasmin
%% programming language and does not support the compilation of
%% cryptographic implementations in C or C++.
%% %
%% Finally, Jasmin only support one target machine language, x86.
\citeauthor{besson_information-flow_2019} present the notion of
\ac{IFL} in compiler optimizations that guarantees that the target
program is not more vulnerable than the source program, i.e.\ that the
transformation does not introduce new
vulnerabilities~\cite{besson_information-flow_2019}.
They use their model to evaluate two passes in CompCert, dead-store
elimination and register allocation.
The evaluated threat model considers observation points at function
boundaries.
In contrast, the \ac{\toolname} backend generates a program secure
against \ac{ROT} and \ac{MRE} leaks at each execution point.
In addition to this, \ac{\toolname} does not guarantee the
preservation of the property but rather the absence of \acp{TBL}.
If that is not possible, the model is unsatisfiable and \ac{\toolname}
fails to generate a program. %\romy{Read this again}
The latter outcome has not appeared in our experiments\footnote{ There
  were unsatisfiable instances due to associativity-related \acp{VBL}
  when using aggressive high-level compiler optimizations (O1, O2, and
  O3)} but there is no guarantee that it will not happen.
For remedying this problem, one may try to activate a pass in
\ac{\toolname} that introduces additional copies of masked values,
deactivate some high-level optimizations, and/or
deactivate the \ac{ROT} or \ac{MRE} constraints.

A recent approach~\cite{vu_secure_2020,vu_reconciling_2021-1}
generates high-quality code that deals with \acp{VBL}, \ac{FI}, and
\ac{TSC} attacks.
To achieve this, \citeauthor{vu_secure_2020}~\cite{vu_secure_2020}
introduce the concept of \textit{opaque observations} that disallows
the compiler to remove security mitigations or rearrange operands in
instructions, such as masking instructions.
In their later work~\cite{vu_reconciling_2021-1}, they improve the
performance of their optimizing compiler by reducing the requirement
for serialization.
To achieve this, they require source-code annotation that may be
challenging for non-trivial programs~\cite{vu_reconciling_2021-1}.
Our approach, \ac{\toolname}, considers \acp{TBL} and, thus, is
complementary to the work of
\citeauthor{vu_secure_2020}~\cite{vu_secure_2020}.
We believe that the combination of \ac{\toolname} with the approach by
\citeauthor{vu_secure_2020} would improve the efficiency of the
generated code.
We leave the adaptation of their methodology in our the front- and
middle-end of \ac{\toolname} as future work.

\subsubsection*{Combinatorial Compiler Approaches}
Compiler backend optimizations, like instruction selection,
instruction scheduling, and register allocation are known to be
hard combinatorial problems.
Hence, solving such problems completely does not scale for large sizes.
Therefore, popular compilers, like GCC~\cite{stallman2009using} and
LLVM~\cite{lattner_llvm_2004}, use heuristics that throughout the
years have proved to improve program performance.
However, these heuristics do not guarantee finding the optimal
solution to these backend optimizations.

For critical code and code aimed for compiler-demanding architectures,
combinatorial methods may find an optimized version of the code that
leads to reduced power consumption and/or high performance benefits.
%% may require the use of  to model the backend optimizations as combinatorial problems
%% and find the model optimal solution that corresponds to the optimized
%% code.
%
Different works~\cite{lozano_survey_2019,lozano_combinatorial_2019,
  eriksson_integrated_2012,gebotys_efficient_1997} aim to optimize
critical code at different levels, like
loops~\cite{eriksson_integrated_2012},
locally~\cite{gebotys_efficient_1997} or at function
level~\cite{lozano_combinatorial_2019}.
The optimization goals range from execution time, code size, or
estimated energy
consumption~\cite{eriksson_integrated_2012,lozano_combinatorial_2019,gebotys_efficient_1997}.
The main drawback of these approaches is
scalability~\cite{lozano_survey_2019}.
However, a recent work, Unison~\cite{lozano_combinatorial_2019},
allows the optimization of functions of up to almost 1000
instructions.

A different combinatorial approach for generating optimal
program code is superoptimization~\cite{fraser_compact_1979}.
Given a code sequence, superoptimization approaches attempt to
find an equivalent code sequence that reduces the overall execution
time and is provably equivalent to the initial code.
Souper~\cite{souper}, a state-of-the-art superoptimization
approach, performs middle-end optimizations to LLVM \ac{IR} code.
Middle-end optimizations typically do not take decisions on the
register allocation or the instruction scheduling.
Instead, they enable algorithmic-level code optimizations.
Crow~\cite{cabrera2021crow} is an approach based on Souper that
performs software diversification as a security mitigation approach.
\citeauthor{eldib_synthesis_2014-2}~\cite{eldib_synthesis_2014-2}
propose a high-level program synthesis approach to automatically
generate masked implementations.
However, the backend compiler transformations that follow these
middle-end transformations, may introduce \acp{TBL}.

To summarize, many combinatorial compiler backend techniques allow
low-level code optimization but, to our knowledge, none of them
considers the preservation of security properties against \acp{TBL}.

%% The verification of the constant-time free register allocation
%% occurs \textit{a posteriori}, i.e., the non-verified pass produces
%% a solution that is then checked by a verifier.
%% %
%% However, Jasmin is 
%% Jasmin, 
%% improved 
%% \romy{
%%   Jasmin \cite{almeida_jasmin_2017},
%%   1) (+) a verified compiler
%%   2) (-) mostly concerns constant-time
%%   3) (-) currently support 1 arch
%%   4) (-) needs to be written in an high-level assembly language
%%          cannot compile already existent code
%%   Information Flow in opt \cite{besson_information-flow_2019}:
%%   %% Souper \cite{souper},
%%   Crow \cite{cabrera2021crow} use of super for diversification?,
%% }
%

\begin{table}
  \centering
  \begin{tabular}{|l|l|c|c|}
    \hline
    Pub. & Mitigation & Target & Processor\\\hline
    %Papagiannopoulos et al.\
    \cite{papagiannopoulos_mind_2017}   &\acs{ROT}, \acs{MOT}, \acs{MRE}, \acs{RNL}  &AVR     & ATMega163\\
    \cite{wang_mitigating_2019-1}       &\acs{ROT}                                  &\faast    & \faast \\
    \cite{athanasiou_automatic_2020-1}  &\acs{ROT}                                  &ARM     & ARM Cortex-M3\\
    \cite{shelton_rosita_2021-1}        &\acs{ROT}, \acs{MOT}, \acs{MRE}, \acs{IPI}, \acs{OT}  & ARM & ARM Cortex-M0 \\
    \hline
    \toolname                          &\acs{ROT}, \acs{MRE}                        &\faast    & \faast\\
    \hline
  \end{tabular}
  \caption{\label{tbl:psc}\ac{TBL}-aware approaches}
\end{table}

\subsubsection*{Code Hardening Against Power Side-Channel Attacks}

There is a number of approaches that deal with different types of
\ac{TBL}-related \acp{PSC}.
Table~\ref{tbl:psc} shows the mitigation approaches against \acp{TBL}.
For each of the related works, Table~\ref{tbl:psc}, presents the
leakage types each of them mitigates (Mitigation), the target
architecture (Target), and the target processor (Processor).
In the last two columns \faast denotes that these approaches may target
multiple architectures and processors.

\citeauthor{papagiannopoulos_mind_2017}~\cite{papagiannopoulos_mind_2017}
perform experiments to identify possible sources of leakage in binary
AVR code on a ATMega163.
They identify sources of leakage including \ac{ROT}, \ac{MOT}, which
occurs when overwriting a value in memory, \ac{MRE}, which occurs
when overwriting a value in the memory bus, and \ac{RNL}, which
occurs when the values of neighboring registers interact with each
other~\cite{papagiannopoulos_mind_2017}.
\citeauthor{papagiannopoulos_mind_2017}~\cite{papagiannopoulos_mind_2017}
observe that \ac{ROT} and \ac{MRE} leakages may be exploited with a
small number of runs, 500, whereas \ac{MOT} requites much more (40K).
Rosita~\cite{shelton_rosita_2021-1} is a recent approach to mitigate
transitional effects that may lead to power side-channel attacks using
an emulation-based technique.
Rosita performs an iterative process to identify power leakages in
software implementations for ARM Cortex M0 and identifies transitional
effects due to \ac{ROT}, \ac{MOT}, \ac{MRE}, \ac{IPI}, and \ac{OT}.
\ac{IPI} occurs when pairs of instructions interact with each other
and \ac{OT} corresponds to interactions between data of different
instructions.
The mitigation introduces a performance overhead of 21\% to 64\%.
In comparisson, \ac{\toolname} is a generic compiler-based approach
that may be applied to multiple hardware architectures and introduces
smaller overhead.
However, a direct comparison would be unfair because Rosita mitigates
more leakage sources.
%
%% \ac{\toolname} mitigates \ac{ROT} and \ac{MRE} leakage sources.
%% \roberto{in compilation or code quality?}

\citeauthor{wang_mitigating_2019-1}~\cite{wang_mitigating_2019-1} uses
a rule-based system~\cite{gao_verifying_2019-1,wang_mitigating_2019-1}
to identify leaks in a masked implementation and perform local
register allocation and instruction selection transformations to
mitigate these leaks in LLVM.
They identify transitional effects due to register reuse, \ac{ROT}.
Their approach is scalable and the mitigation introduces small
performance overhead compared to non-optimized code.
However, they depend on a non-optimized compilation in order to
preserve the security properties of the high-level program, which
leads to code generation that is secure against \ac{ROT} but not
optimized.
\citeauthor{athanasiou_automatic_2020-1}~\cite{athanasiou_automatic_2020-1}
use the same rule-based system to mitigate \ac{ROT} leakages on binary
ARM code targeting the ARM Cortex M3 processor.
They are able to reduce the number of potentially vulnerable register
pairs given the instruction order.
\citeauthor{athanasiou_automatic_2020-1} confirm that aggressive
compiler optimization passes introduce \acp{VBL}.
\ac{\toolname} uses a rule-based system but models a constraint
model that is able to generate optimized code that is secure.

Other approaches perform mitigations at whole-system design
time~\cite{sijacic_towards_2018,gigerl_coco_2021-1}.
The availability of open hardware architectures and, more
specifically, RISC-V, has enabled approaches, such as Coco, which
apply software-hardware co-design techniques to mitigate power
side-channel attacks~\cite{gigerl_coco_2021-1}.

In summary, there are compiler-based and binary rewriting approaches
to mitigate \acp{TBL} but all these approaches perform local
transformations that introduce performance overhead.
Instead \ac{\toolname} trades quality for compilation time and is
suitable for performance critical and vulnerable cryptographic
functions.

\section{Limitations}
\label{sec:limitations}
%% \roberto{sorry to insist, but I would rename this section to ``Limitations''. In
%%   a ``discussion'', I expect something more general, like pros and cons, how the
%%   work fits on a larger scale, etc.}
%
%% \roberto{proposal: rename section to ``Limitations"}
%
This paper proposes an architecture-agnostic method to generate high
quality code against register-reuse and memory-bus transitional
effects.
We aim specifically at small-size embedded devices that have a
predictable cost model and implement single-issued, non-speculative
architectures.
Our approach has clear scalability issues, however, we plan to
investigate its use in non-linearized functions.

Secondly, our approach is limited to two optimizations, namely
register allocation and instruction scheduling.
Other backend optimizations, such as instruction selection may be
beneficial for removing \ac{HD} leakages for CISC architectures like
x86.
Another useful optimization for mitigating optimized implementations
may be expression reassociation (\texttt{-reassociate} in LLVM).

\section{Conclusion and Future Work}

This paper proposes a constraint-based compiler backend to generate
code that is both optimized and secure against power side-channel
attacks.
We prove that the generated code is secure according to a non-trivial
leakage model, and show that our approach achieves high code
improvement against non-optimized approaches ranging from \lowestsu to
a speedup of \largestsu for two embedded architectures, \mips and ARM
Cortex M0.
At the same time, our approach introduces a maximum overhead of
\largestoh from the optimal code.
This comes at the expense of increased compilation time and reduced
scalability.

There are several future directions for our work.
We are planning to work on extending the type-inference algorithm to
include function calls and loops.
Moreover, by improving the accuracy of the hardware model of
\ac{\toolname} to model precisely a specific device, we will be able
to improve the leakage model and compare our approach to approaches
like Rosita~\cite{shelton_rosita_2021-1}.
%
%% A second direction of future work is to improve the scalability of
%% \ac{\toolname} through decomposition.
%% %
%% The challenge here is to find appropriate program points to split
%% linearized code to different blocks.
%
%% We also plan to extend our approach to other mitigations such as
%% timing side channels of constant resource
%% programs~\cite{barthe_secure_2021}.
%
Finally, we believe that combining our approach with optimizing
high-level approaches~\cite{vu_secure_2020,vu_reconciling_2021-1} may
further improve the quality of the generated code.

\section*{Acknowledgment}
We would like to thank Jingbo Wang for providing support for the
FSE19 tool and Amir M. Ahmadian for the fruitful discussions and
his significant feedback on the paper.
Finally, we would like to thank Oscar Eriksson for proof reading the
paper.

\bibliographystyle{IEEEtranN}
\bibliography{IEEEabrv,bibliography}

%\begin{comment}
\appendices
\section{Type Inference Rules}
\label{ap:tinf}

\begin{figure*}
  \centering
  \footnotesize
\[
\begin{array}{ccc}
  \infr[RAND]{dom(t) \neq \varnothing}
     {\Gamma \vdash t: Rand}
&
     \infr[PUB$_1$]{supp(t) \cap IN_{sec} = \varnothing \andrule dom(t) = \varnothing}
          {\Gamma \vdash t:Pub }
               &
\infr[PUB$_{2}$]{\Gamma \vdash t_0 :Pub \andrule \Gamma \vdash t_1 : Pub \\\andrule  supp(t_0) \cap supp(t_1) = \varnothing }
     {\Gamma \vdash t = t_0 \circledast t_1 : Pub}
     \\\\
     \infr[PUB$_{3}$]{\Gamma \vdash t_0 :Rand \andrule \Gamma \vdash t_1 :Rand \\\andrule  (dom(t_0)\backslash supp(t_1) \neq \varnothing ~\lor \\\andrule dom(t_1)\backslash supp(t_0) \neq \varnothing)}
     {\Gamma \vdash t_0 \circ t_1 : Pub}
     &
     \infr[PUB$_{4}$]{\Gamma \vdash t_0: Rand \andrule \Gamma \vdash t_1 :Rand \\ (dom(t_0) \backslash dom(t_1) \neq \varnothing~\lor\\\andrule dom(t_1) \backslash dom(t_0) \neq \varnothing)}
          {\Gamma \vdash t_0\,\odot\,t_1 : Pub}
     &
          \infr[PUB$_{5}$]{  \hfill i\in\{0,1\} \andrule j = 1-i\andrule \Gamma \vdash t_i :Rand \hfill \\ \andrule  dom(t_i)\backslash supp(t_j) = \varnothing \andrule \\ dom(t_i)= dom(t_j) \andrule  supp(t_i)= supp(t_j) }
               {\Gamma \vdash t_0 \circledast t_1 : Pub}
          \\\\
\infr[PUB$_{6}$]{i\in\{0,1\} \andrule j = 1-i \\\andrule \Gamma \vdash t_i :Rand \andrule \Gamma \vdash t_j :Pub \\ \andrule  dom(t_i)\backslash supp(t_j) \neq \varnothing}
     {\Gamma \vdash t_0\,\odot\,t_1 : Pub}
     &
     \infr[PUB$_{7}$]{i\in\{0,1\} \andrule j = 1-i\\\andrule \Gamma \vdash t_i :Pub \andrule \Gamma \vdash t_j : Rand \\ \andrule  supp(t_i) \cap supp(t_j) = \varnothing }
          {\Gamma \vdash t_0 \circ t_1 : Pub}
          &
     \infr[PUB$_{8}$]{(t_0 = t_1\,\odot\,t_2 \,\lor\, t_1 = t_0\,\odot\,t_2 ) \\ \Gamma \vdash t_{0,1}: T_1 \andrule \Gamma \vdash t_2: T_2 \\ T_1 \neq Sec \land T_2 \neq Sec}
     {\Gamma \vdash t_0\,\oplus\,t_1 : Pub}          
     \\\\
     \infr[NEST$_{1}$]{ ((t_0 = t_1\,\oplus\,t_2)\, \lor\, (t_1 = t_0\,\oplus\,t_2) )\\
       \hfill\Gamma \vdash t_2: T \hfill}
     {\Gamma \vdash t_0\,\oplus\,t_1 : T}
     &
     \infr[NEST$_{2}$]{((t_0 = t_1\,\lor\,t_2) \,\lor\, (t_1 = t_0\,\lor\,t_2) )\\
       \hfill \Gamma \vdash \neg t_{0,1}\,\land\,t_2: T \hfill}
     {\Gamma \vdash t_0\,\oplus\,t_1 : T}
     &
     \infr[NEST$_{3}$]{(t_0 = t_1\,\land\,t_2 \land t_1 = t_0\,\land\,t_2 )\\
       \hfill\Gamma \vdash t_{0,1}\,\land\,\neg t_2: T\hfill}
          {\Gamma \vdash t_0\,\oplus\,t_1 : T}
          \\\\
     \infr[DISTR$_0$]{ 
       \Gamma \vdash t_{0} \,\odot\,(t_{1} \,\oplus\,t_{2}): T}
     {\Gamma \vdash (t_{0}\,\odot\,t_{1})\,\oplus\,(t_{0}\,\odot\,t_{2}) : T}
     %% \infr[DISTR]{ 
     %%   (t_{00} = t_{10} \,\lor\, t_{00} = t_{11} \,\lor\, t_{01} = t_{10} \,\lor\, t_{01} = t_{11})\\
     %%   \hfill\Gamma \vdash t_2: T \hfill}
     %% {\Gamma \vdash (t_{00}\,\odot\,t_{01})\,\oplus\,(t_{10}\,\odot\,t_{11}) : T}
     &
          \infr[DISTR$_1$]{ 
       \Gamma \vdash t_{0} \,\odot\,(t_{1} \,\oplus\,t_{2}): T}
     {\Gamma \vdash (t_{0}\,\odot\,t_{1})\,\oplus\,(t_{2}\,\odot\,t_{0}) : T}
     &
     \infr[DISTR$_2$]{ 
       \Gamma \vdash t_{1} \,\odot\,(t_{0} \,\oplus\,t_{2}): T}
     {\Gamma \vdash (t_{0}\,\odot\,t_{1})\,\oplus\,(t_{1}\,\odot\,t_{2}) : T}
     \\\\
     \infr[DISTR$_3$]{ 
       \Gamma \vdash t_{1} \,\odot\,(t_{0} \,\oplus\,t_{2}): T}
     {\Gamma \vdash (t_{0}\,\odot\,t_{1})\,\oplus\,(t_{2}\,\odot\,t_{1}) : T}
     &
     \infr[PUB$_{9}$]{\hfill\Gamma \vdash t_0 :Pub \andrule \Gamma \vdash t_1 : Pub \hfill\\
       \andrule  supp(t_0) \cap supp(t_1) \cap IN_{rand} = \varnothing }
     {\Gamma \vdash t = t_0 \oplus t_1 : Pub}

     &
\end{array}
\]
\caption{\label{fig:typsys} Type inference for power side channels in
  \ac{\toolname}~\cite{wang_mitigating_2019-1}; $\oplus$ denotes the
  exclusive OR operation, $\odot$ denotes the multiplication in a
  finite field, $\circ$ denotes any other operations apart from
  $\odot$ and $\oplus$, and finally, $\circledast$ denotes any operation.}
\end{figure*}

The security analysis of \ac{\toolname} requires the type annotation of
program variables and variables generated by the transformations of
the underlying constraint-based compiler backend.
We have implemented the type inference algorithm by
\citeauthor{wang_mitigating_2019-1}~\cite{wang_mitigating_2019-1} due
to its scalability compared with other approaches like symbolic
execution~\cite{bayrak_sleuth_2013}.
This section describes the type inference algorithm starting with the
definition of auxiliary functions.
Although \ac{\toolname} uses multiple equivalent temporary (copy)
values for each operation operand (see Figure~\ref{lst:unimasked}),
definitions use a single temporary value $t$.
In reality, we unify these equivalent temporaries because they are
semantically equal, as they are just copies of the original
program variables.
In the following definition, the parts in \textbf{bold} denote the
extensions to the original type-inference
algorithm~\cite{wang_mitigating_2019-1}.

The auxiliary function \textit{xor}, returns true if an expression only
consists of exclusive OR operations.
This function improves the precision of the type inference algorithm,
when multiple exclusive OR operations remove the dependence on a
secret value.
The recursive definition of \textit{xor} is as follows:

\begin{tcolorbox}[colback=\colbackts,colframe=\colframets, top=-5pt, bottom = 0pt,
  right = 0pt, left = 0pt]%,title=My nice heading]
\boldmath\begin{equation*}
xor(t_0) =
  \begin{cases}
    \textbf{\texttt{true}} & \text{ if }t_0\in IN\\
    xor(t_1) & \text{ if }t_0 = uop(t_1)\\
    xor(t_1) \land xor(t_2) &\text{ if }t_0 = \oplus(t_1,t_2)\\
    \textbf{\texttt{false}} &\text{ if }t_0 = bop(t_1,t_2),\\
         &~~~~bop\neq \oplus
  \end{cases}
\end{equation*}
\end{tcolorbox}

The auxiliary function \textit{supp}~\cite{wang_mitigating_2019-1} returns the
support of each expression.
That is, all the variables that are syntactically present in the
expression.
We add two cases for \textit{supp}, where some of syntactically present
values are removed in the case of a simplification.
This improves the precision of the analysis, because the type
inference algorithm uses \textit{supp} to decide on the type of a
temporary variable.
The recursive definition of \textit{supp} is:

\begin{tcolorbox}[colback=\colbackts,colframe=\colframets, top=-5pt, bottom = 0pt,
  right = 0pt, left = 0pt]%,title=My nice heading]
\begin{align*}
&supp(t_0) =\\
&  \begin{cases}
    \{t_0\} & \text{ if }t_0\in IN\\
    supp(t_1) & \text{ if }t_0 = uop(t_1)\\
%%    supp(t_1) \cup supp(t_2) & \text{ if }t_0 = bop(t_1,t_2)
    \boldsymbol{(supp(t_1) \cup supp(t_2))\backslash} &\boldsymbol{\text{ if }t_0 = \oplus(t_1,t_2) \land} \\
    \boldsymbol{(supp(t_1) \cap supp(t_2))}  &~~~~~~~~~~ \boldsymbol{xor(t_0)}\\
    \boldsymbol{supp(t_2)} \!\!\!\!\!\!&\boldmath{\text{ if }t_0 = \oplus(t_1,\oplus(t_1,t_2))}\\
    supp(t_1) \cup supp(t_2) & \text{ if }t_0 = bop(t_1,t_2)
  \end{cases}
\end{align*}
\end{tcolorbox}

The definitions of \textit{unq} and \textit{dom} are the same
as the original definitions by \citeauthor{wang_mitigating_2019-1}.
We define them here for completeness.

Auxiliary function \textit{unq}~\cite{wang_mitigating_2019-1} returns
the random input variables that appear only once in the expression.
This means that if we have a binary operator \textit{bop}, with two
operands $t_1$ and $t_2$ then, if both operands are randomized with
the same random value, then this random value cannot randomize the
expression $t_0$.
The recursive definition of \textit{unq} is:

\begin{tcolorbox}[colback=\colbackts,colframe=\colframets, top=-5pt, bottom = 0pt,
    right = 0pt, left = 0pt]%,title=My nice heading]
\begin{equation*}
unq(t_0) =
  \begin{cases}
    \{t_0\} & \text{ if }t_0\in IN_{rand}\\
    \{\} & \text{ if }t_0\in IN\backslash IN_{rand}\\
    unq(t_1) & \text{ if }t_0 = uop(t_1)\\
    (unq(t_1) \cup unq(t_2))\backslash \!\!\!\!\!\!&\\
    (supp(t_1) \cap supp(t_2)) \!\!\!\!\!\! & \text{ if }t_0 = bop(t_1,t_2)
  \end{cases}
\end{equation*}
\end{tcolorbox}

The last auxiliary function is \textit{dom}~\cite{wang_mitigating_2019-1}.
For each temporary variable, \textit{dom} returns the random input variables
that are xor:ed with that value.
The recursive definition of \textit{dom} is:

\begin{tcolorbox}[colback=\colbackts,colframe=\colframets, top=-5pt, bottom = 0pt,
  right = 0pt, left = 0pt]%,title=My nice heading]
\begin{align*}
& dom(t_0) = \\
&  \begin{cases}
     \{t_0\} & \text{ if }t_0\in IN_{rand}\\
     \{\} & \text{ if }t_0\in IN\backslash IN_{rand}\\
     dom(t_1) & \text{ if }t_0 = uop(t_1)\\
     (dom(t_1) \cup dom(t_2))\!\!\!\!\!&\\
     ~~~~\cap\, unq(t_0) &\text{ if }t_0 = \oplus(t_1,t_2)\\
     \{\} &\text{ if }t_0 = bop(t_1,t_2) \land bop\neq \oplus
   \end{cases}
\end{align*}
\end{tcolorbox}

Finally, Figure~\ref{fig:typsys} presents the type system.
Rules \texttt{RAND} and \texttt{PUB$_1$} to \texttt{PUB$_8$} are
described by
\citeauthor{wang_mitigating_2019-1}~\cite{wang_mitigating_2019-1} and
the rest of the rules are discussed in the same paper.
Here, for space reasons, we have abbreviated \texttt{Random} to
\textit{Rand}, \texttt{Public} to \textit{Pub}, and \texttt{Secret} to
\textit{Sec}.
In particular, the first two rules are the basic rules, i.e.\ 1) if
\textit{dom} for an expression contains a value, then, this temporary
has type \textit{Rand}, and 2) if the type is not \textit{Rand} and
the expression does not depend on secret values, then the expression
has type \textit{Pub}.
The rest of the rules improve the precision of the analysis.
In particular, rules DISTR$_0$ to DISTR$_3$ are new rules
that do not appear in \citeauthor{wang_mitigating_2019-1}.

\section{Security Proof}
\label{ap:secproof}
We assume that the type-inference
algorithm~\cite{wang_mitigating_2019-1} is conservative and sound:
if $type(t) = Rand$, then $t$ follows a uniform
random distribution;
if $type(t) = Pub$, then $t$ follows a secret-independent
distribution (might also be uniform random distribution); and
if $type(t) = Sec$, then $t$ may be secret
dependent.

\ac{\toolname} generates a solution to the constraint model, which we
represent as an ordered sequence of instructions, $P = \{i_0, ...,
i_n\}$.
This means that instruction $i_j$ is executed before instruction $i_k$
for $j < k$.

To verify whether the generated program leaks secret information
according to our leakage model
(Equations~\ref{eq:leakage1}-\ref{eq:leakage4}), we give a proof of
Theorem~\ref{th:proof} using structural induction on a mitigated
program, $P$.
%% We shall give a fairly detailed proof of the lemma using structural induc-
%% tion on the arithmetic expressions.
%% structural proof on the program, $P$.
%
We start from the last instruction because preceding instructions are
able to hide the secret values.
%%%%%%%%%%%%%%%%%%%%%%%%%%%%%%%% New Proof Style %%%%%%%%%%%%%%%%%%%%%%%%%%%%%%
\begin{proofItemize}
\item[] \textbf{Case 1} Assume $P = t \leftarrow e$.
  
  From the leakage model (Equation~\ref{eq:leakage3}), we have
  $L(P(IN)) = \{HW(e \oplus r_{IN})\}$, where $r(t) = r$).

  \begin{proofItemize}
                \item[] \textbf{Case 1.a} Assume $type(r_{IN}) = Pub$.
                
                This means that the input is a constant value.
                \begin{proofItemize}
                    \item[] \textbf{Case 1.a.i} Assume $type(e) \in \{Rand, Pub\}$.
                
		      Because $type(e) \in \{Rand, Pub\}$, the distribution of $e$
		      is either random (uniformly distributed) or public.
		      This means that the distribution is not dependent on
		      the secret value.
		      Thus, Definition~\ref{def:security_condition} is satisfied.
		      
		    \item[] \textbf{Case 1.a.ii} Assume $type(e) = Sec$.
		      
		      From the definition of \textit{Spairs} (Equation~\ref{eq:spairs}),
		      $type(e) = type(t) = Sec \implies \exists (t_i, ts) \in Spairs.~ t_i =
		      t ~\land~ ts = [t' | t'\in Temps ~\land~type(t') = Rand ~\land~
		        type(t'\oplus t) = Rand]$.
		      In this case, we have a pair $(t, ts)$, but the first
		      constraint in Section~\ref{ssec:sconstraint} is not
                      satisfied because $\nexists t_r \in ts~.~ \texttt{subseq}(t_r, t)$
                      (Theorem~\ref{th:instrseq}).
		      So, $P$ is not a valid program.
              \end{proofItemize}

              \item[] \textbf{Case 1.b} Assume $type(r_{IN}) = Rand$.

              \begin{proofItemize}
                    \item[] \textbf{Case 1.b.i} Assume $type(e) \in \{Rand, Pub\}$.

                      \begin{proofItemize}
                      \item[] \textbf{Case 1.b.i.$\alpha$} Assume
                        $type(e \oplus r_{IN}) \in \{Rand, Pub\}$.
                
		        Because $type(e \oplus r_{IN}) \in \{Rand,
                        Pub\}$, the distribution is either random
                        (uniformly distributed) or public.
		        This means that the distribution is not dependent on
		        the secret value.
		        Thus, Definition~\ref{def:security_condition} is satisfied.
		        
		      \item[] \textbf{Case 1.b.i.$\beta$} Assume
                        $type(e \oplus r_{IN}) = Sec$.

                        From the hypotheses in Case 1.b and Case 1.b.i
                        and the definition of $Rpairs$, we have that
                        $(t(r_{IN}), t) \in Rpairs$.
                        This means that the constraint in Section~\ref{ssec:rconstraint}
                        is not satisfied because we have that
                        $\texttt{subseq}(t(r_{IN}), t)$.
		        Hence, $P$ is not a valid program.
                      \end{proofItemize}

		    \item[] \textbf{Case 1.b.ii} Assume $type(e) = Sec$.
		      
		      From the definition of \textit{Spairs} (Equation~\ref{eq:spairs}),
		      $type(e) = type(t) = Sec \implies \exists (t_i, ts) \in Spairs.~ t_i =
		      t ~\land~ ts = [t' | t'\in Temps ~\land~type(t') = Rand ~\land~
		        type(t'\oplus t) = Rand]$.
                      \begin{proofItemize}
                      \item[] \textbf{Case 1.b.ii.$\bm{\alpha}$} Assume
                        $type(e \oplus r_{IN}) \in \{Rand, Pub\}$.
                
		        Because $type(e \oplus r_{IN}) \in \{Rand,
                        Pub\}$, the distribution is either random
                        (uniformly distributed) or public.
		        This means that the distribution is not dependent on
		        the secret value.
		        Thus, Definition~\ref{def:security_condition} is satisfied.
		        
		      \item[] \textbf{Case 1.b.ii.$\bm{\beta}$} Assume
                        $type(e \oplus r_{IN}) = Sec$.

                        From Theorem~\ref{th:instrseq}, we have that
                        $\texttt{subseq}(t(r_{IN}), t)$.
                        Also, there is no other $t' \in Temps$ such
                        that $\texttt{subseq}(t', t)$, i.e.\ $\nexists
                        t' \neq t(r_{IN}) ~.~ \texttt{subseq}(t', t)$.
                        From  the first constraint
                        in Section~\ref{ssec:sconstraint}, we have
                        that $\exists t' \in ts ~.~ \texttt{subseq}(t', t)$.
                        Which means that $t' = t(r_{IN})$.
                        However, 
                        if $t(r_{IN})\in ts$ then $type(e \oplus
                        r_{IN}) = Rand$ (Equation~\ref{eq:spairs}),
                        which is not true.
		        Hence, $P$ is not a valid program.
                      \end{proofItemize}
                \end{proofItemize}
              \item[] \textbf{Case 1.c} Assume $type(r_{IN}) = Sec$.

              \begin{proofItemize}
                   \item[] \textbf{Case 1.c.i} Assume
                     $type(e \oplus r_{IN}) \in \{Rand, Pub\}$.

		     Because $type(e \oplus r_{IN}) \in \{Rand,
                     Pub\}$, the distribution is either random
                     (uniformly distributed) or public.
		     This means that the distribution is not dependent on
		     the secret value.
		     Thus, Definition~\ref{def:security_condition} is satisfied.
		        
		    \item[] \textbf{Case 1.c.ii} Assume $type(e \oplus r_{IN}) = Sec$.
		      
		      From the definition of \textit{Spairs}
                      (Equation~\ref{eq:spairs}), $type(r_{IN}) =
                      type(t(r_{IN})) = Sec \implies \exists (t_i, ts)
                      \in Spairs.~ t_i = t ~\land~ ts = [t' | t'\in
                        Temps ~\land~type(t') = Rand ~\land~
                        type(t'\oplus t) = Rand]$.
                      From Theorem~\ref{th:instrseq}, we have that
                      $\texttt{subseq}(t(r_{IN}), t)$.
                      Also, there is no other $t' \in Temps$ such
                      that $\texttt{subseq}(t(r_{IN}, t)$, i.e.\ $\nexists
                      t' \neq t ~.~ \texttt{subseq}(t, t')$.
                      From  the second constraint
                      in Section~\ref{ssec:sconstraint}, we have
                      that $\exists t' \in ts ~.~ \texttt{subseq}(t(r_{IN}, t'))$.
                      Which means that $t' = t$.
                      However, if $t'\in ts$ then $type(t \oplus
                      t(r_{IN})) = Rand$ (Equation~\ref{eq:spairs}),
                      which is not valid from hypothesis (Case
                      1.c.ii).
		      Hence, $P$ is not a valid program.

                \end{proofItemize}

  \end{proofItemize}

	\item[] \textbf{Case 2} Assume $P = mem(e_a,e)$.
	\begin{proofItemize}
		\item[] \textbf{Case 2.a} Assume $type(e) \in \{Rand,Pub\}$
		From the leakage model (Equation~\ref{eq:leakage4}), we have
		$L(P(IN)) = \{HW(e)\}$.
		Because $type(e) \in \{Rand, Pub\}$, the distribution of $e$
		is either random (uniformly distributed) or public, i.e. a constant
		value.
		This means that the distribution is not dependent on
		the secret value.
		Thus, Definition~\ref{def:security_condition} is satisfied.
		
		\item[] \textbf{Case 2.b} Assume $type(e) = Sec$.
		
		From the definition of \textit{Mspairs} (Equation~\ref{eq:mspairs}),
		$type(e) = Sec \implies \exists (o_i, os) \in Mspairs.~ tm(o_i) = e 
		~\land~ os = [o' | o'\in \mathit{MemOperations} ~\land~type(tm(o')) = Rand ~\land~
		type(tm(o')\oplus tm(o)) = Rand]$.
		In this case we have a pair $(o, \varnothing)$, and thus,
		the constraint in Section~\ref{ssec:msconstraint} is not satisfied,
		because $\nexists o_i \in \varnothing$.
		So, $P$ is not a valid program.
	\end{proofItemize}

	\item[] \textbf{Case 3} Assume $P = P' ; t \leftarrow e$.
	\begin{proofItemize}
		\item[] \textbf{Case 3.a} Assume $type(e) = Sec$.

		From the definition of \textit{Spairs} (Equation~\ref{eq:spairs}),
		$type(e) = type(t) = Sec \implies \exists (t_i, ts) \in Spairs.~ t_i =
		t ~\land~ ts = [t' | t'\in Temps ~\land~type(t') = Rand ~\land~
		type(t'\oplus t) = Rand]$.

		From the Spairs constraint in Section~\ref{ssec:sconstraint}, we
		have that $\exists t_r \in ts.~l(t) \implies l(t_r)~\land~
		\texttt{subseq}(t_r, t)$.
		From Theorem~\ref{th:instrseq}, we have $\texttt{subseq}(t_r, t)
		\implies P = P''; t_r \leftarrow e_r; P'''; t \leftarrow e ~\land~
		r(t) = r(t_r) ~\land~ \forall i \leftarrow P'''.~ i= t' \leftarrow e'
		~\land~ r(t') \neq r(t)$.
		According to the leakage model (Equations~\ref{eq:leakage1}),
		$L(P) = L(P''; t_r \leftarrow e_r; P''') \cup \{ HW(t_r \oplus t)\} $.
		Because $t_r\in ts$, we have that $type(t_r \oplus t) = Rand$.
		This means that $t_r \oplus t$ has a uniform random distribution, and,
		thus, $HW(t_r \oplus t)$ does not leak.
		From the induction hypothesis, $\sum_{l\in L(P'(IN))} \mathbb{E}[l] =
		\sum_{l \in L(P'(IN'))}\mathbb{E}[l]$ and $\sum_{l\in L(P'(IN))}
		var[l] = \sum_{l \in L(P'(IN'))}var[l]$.
		Thus, $\sum_{l\in L(P(IN))} \mathbb{E}[l] = \sum_{l\in L(P(IN))} \mathbb{E}[l] + HW(t_r \oplus t) =
		\sum_{l\in L(P(IN'))} \mathbb{E}[l] + HW(t_r \oplus t) = \sum_{l\in L(P(IN'))} \mathbb{E}[l]$.
		Same is true for $var$.
		Thus, Definition~\ref{def:security_condition} is satisfied.
		
		\item[] \textbf{Case 3.b} Assume $type(e) \in \{Rand, Pub\}$.
		\begin{proofItemize}
			\item[] \textbf{Case 3.b.i} Assume $\exists i \in P'.~ i = t' \leftarrow e'~\land~ r(t)= r(t')$.
			
			Of the temporaries assigned to the same register, we select the
			temporary that is scheduled last before $t$, i.e.\ $P = P''; t_r
			\leftarrow e_r; P'''; t \leftarrow e ~\land~ \forall i \leftarrow
			P'''.~ i= t' \leftarrow e' ~\land~ r(t') \neq r(t)$
			\begin{proofItemize}
				\item[] \textbf{Case 3.b.i.$\bm{\alpha}$} Assume $type(t \oplus t') \in \{Rand, Pub\}$.
				
				In this case, the leakage model is $L(P) = L(P''; t_r \leftarrow
				e_r; P''') \cup \{ HW(t \oplus t')\} $.
				Due to the initial assumption $type(t \oplus t') \in \{R, P\}$, the
				distribution of the leakage is either randomly distributed or public,
				i.e. it does not reveal secret information.
				From the induction hypothesis, $\sum_{l\in L(P'(IN))} \mathbb{E}[l] =
				\sum_{l \in L(P'(IN'))}\mathbb{E}[l]$ and $\sum_{l\in L(P'(IN))}
				var[l] = \sum_{l \in L(P'(IN'))}var[l]$.
				Thus, $\sum_{l\in L(P(IN))} \mathbb{E}[l] = \sum_{l\in L(P(IN))} \mathbb{E}[l] + HW(t \oplus t') =
				\sum_{l\in L(P(IN'))} \mathbb{E}[l] + HW(t \oplus t') = \sum_{l\in L(P(IN'))} \mathbb{E}[l]$.
				Same is true for $var$.
				Thus, Definition~\ref{def:security_condition} is satisfied.
				
				\item[] \textbf{Case 3.b.i.$\bm{\beta}$} Assume $type(t \oplus t') = Sec$.
				\begin{proofItemize}
					\item[] \textbf{Case 3.b.i.$\bm{\beta}$.1} Assume $type(t') \in \{Rand,Pub\}$.
					
					From the definition of $Rpairs$ (Equation~\ref{eq:rpairs}),
					$(t,t') \in Rpairs$.
					From the  $Rpairs$ constraint in Section~\ref{sec:constraint_model}, we
					have that $\neg \texttt{subseq}(t,t') \land \neg\texttt{subseq}(t',t)$.
                                        From the definition of \texttt{subseq}, the first term,
                                        $\neg\texttt{subseq}(t,t')$,  is \texttt{true} 
					because $t$ follows $t'$ in the program
					sequence.
					The second constraint $\neg\texttt{subseq}(t',t)$ contradicts with the
					hypothesis in \textit{Case 3.a.i} (Theorem~\ref{th:instrseq}).
					
					\item[] \textbf{Case 3.b.i.$\bm{\beta}$.2} Assume $type(t') = Sec$.
					
					From the definition of $Spairs$ (Equation~\ref{eq:spairs}) we have
					that $\exists (t_i, ts) \in Spairs.~t_i = t'$ with $\forall t_s \in ts
					.~ type(t' \oplus t_s) = Rand$.
					From the $Spairs$ constraint in
					Section~\ref{sec:constraint_model}, $\exists t_r \in ts . ~ l(t')
					\implies l(t_r) ~\land~ \texttt{subseq}(t',t_r)$.
					However, because there is no other assignment to register $r(t)$ in
					$P'''$ (\textit{Case 3.b.i}), we have that $t_r = t$ and because $t_r
					\in ts$, $type(t_r \oplus t') = Rand$.
					But $type(t \oplus t') = Sec$ (\textit{Case 2.b.i}), which is a
					contradiction.
				\end{proofItemize}
			\end{proofItemize}
			
			\item[] \textbf{Case 3.b.ii} Assume $\nexists i \in P'.~ i = t' \leftarrow e'~\land~ r(t)= r(t')$.
			
			Then, the leakage is $L(P) = L(P') \cup \{ HW(e)\}$.
			$HW(e)$ follows either a random distribution or is secret independent.
			From the induction hypothesis, $\sum_{l\in L(P'(IN))} \mathbb{E}[l] =
			\sum_{l \in L(P'(IN'))}\mathbb{E}[l]$ and $\sum_{l\in L(P'(IN))}
			var[l] = \sum_{l \in L(P'(IN'))}var[l]$.
			Thus, $\sum_{l\in L(P(IN))} \mathbb{E}[l] = \sum_{l\in L(P(IN))} \mathbb{E}[l] + HW(e) =
			\sum_{l\in L(P(IN'))} \mathbb{E}[l] + HW(e) = \sum_{l\in L(P(IN'))} \mathbb{E}[l]$.
			Same is true for $var$.
			Thus, Definition~\ref{def:security_condition} is satisfied.
		\end{proofItemize}
	\end{proofItemize}
	
	\item[] \textbf{Case 4} Assume $P = P' ; mem(e, e_i)$.
	\begin{proofItemize}
		\item[] \textbf{Case 4.a} Assume $type(e) = Sec$.
		
		Analogous to \textit{Case 3.a}.
		
		\item[] \textbf{Case 4.b} Assume $type(e) \in \{ Rand, Pub\}$.
		
		Analogous to \textit{Case 3.b}.
	\end{proofItemize} 

\end{proofItemize}

\section{Implied Constraints}
\label{sec:implied}
To improve the solver's ability to find solutions, we add additional
constraints that are logically implied by the imposed constraints.
Implied constraints often improve the solving procedure by reducing
the search space through propagation~\cite{rossi2006handbook}.

The following implied constraint is specifically relevant to ARM
Cortex M0 but also to architectures that use accumulators for many
operations, such as x86 architectures.
This constraint enforces that if a pair of temporaries in
$Rpairs$ belong to the same operation $\texttt{o}$ then the two operands
(destination and source) have to be assigned to different registers or
the operation operands should change.
If the source and destination operands have to be assigned to the same
register (accumulator) then, the operands have to be inverted.
The constraint is as follows:

\begin{tcolorbox}[colback=\colbackcolor,colframe=\colframecolor,
    top=-5pt, bottom = -5pt,
    right = 0pt, left = 0pt]
\begin{lstlisting}[style = modelingstyle]
forall (t$_1$,t$_2$) in $Rpairs$:
  o = def_oper(t$_1$)
  if (o $\in$ user_opers(t$_2$)):
    $\neg$same_reg(t$_1$, t$_2$)
\end{lstlisting}
\end{tcolorbox}

Another implied constraint is related to preassigned operands.
Preassigned operands are given a specific register because
of special hardware architecture properties or calling conventions.
For this, we add an additional implied constraint that guides the
solver to try to schedule a different temporary if the two preassigned
temporaries are not allowed to be subsequent, i.e.\ they belong to
$Rpairs$.
\begin{tcolorbox}[colback=\colbackcolor,colframe=\colframecolor,
    top=-5pt, bottom = -5pt,
    right = 0pt, left = 0pt]
\begin{lstlisting}[style = modelingstyle]
forall (t$_1$,t$_2$) in $Rpairs$:
  if (t$_2$ $\in$ preassign $\land$ t$_1$ $\in$ preassign):
    samereg(t$_1$, t$_2$) $\implies$ (
     (exists t $\in$ Temps: subseq(t1,t) $\lor$
                        subseq(t,t1)) $\land$
     (exists t $\in$ Temps: subseq(t2,t) $\lor$
                        subseq(t,t2)))
\end{lstlisting}
\end{tcolorbox}
%\end{comment}

\end{document}